\documentclass[12pt]{article}

\usepackage{epsfig,latexsym,amsfonts,amsmath,amsthm,amssymb,amsbsy,multirow,slashed,wasysym,textcomp,dsfont,comment,mathtools,cancel,cite,diagbox,datetime,appendix,BOONDOX-calo}
\usepackage{tocloft}
\usepackage{bbold}
\usepackage{sidecap}
\usepackage{adjustbox}
\usepackage{pgfplots}
\usepackage{oplotsymbl}
\usepackage{tikzpagenodes}
\usepackage{adforn}
\usepackage{tikz}
\usetikzlibrary{backgrounds}
\usetikzlibrary{patterns}
\usetikzlibrary{arrows.meta}
\usetikzlibrary{shapes}
\tikzstyle{bag} = [align=center]
\usetikzlibrary{decorations.pathmorphing}
\usepackage[normalem]{ulem}
\usepackage[mathscr]{eucal}
\usepackage[version=3]{mhchem}

\usepackage{hyperref}
\usepackage{subcaption}

 \newcommand{\badat}{\begin{alignedat}}
 \newcommand{\eadat}{\end{alignedat}}
 \newcommand\scalemath[2]{\scalebox{#1}{\mbox{\ensuremath{\displaystyle #2}}}}
 \def\be{\begin{equation}}
\def\ee{\end{equation}}

\def\p{\partial}

\usepackage{MnSymbol}

\usepackage{color}

\newcommand{\pink}[1]{\textcolor{\pink}{#1}}

\definecolor{dblue}{rgb}{0.2,0.50,0.80}

\def\D{\mathcal{D}}

\def\O{\mathcal{O}}

\def\bh{{\bar h}}
\def\bz{{\bar z}}

\def\ba{{\bar a}}
\def\bb{{\bar b}}
\def\bc{{\bar c}}
\def\bd{{\bar d}}

\def\zb{\bar{z}}

\def\bh{{\bar h}}
\def\bz{{\bar z}}

\def\a{\alpha}
\def\b{\beta}
\def\pa{\partial}
\def\D{\Delta}
\def\o{\omega}
\def\d{\delta}
\def\t{\tau}

\numberwithin{equation}{section} 
\pgfplotsset{compat=1.17} 
\begin{document}

 \begin{titlepage}
  \thispagestyle{empty}
  \begin{flushright}
  \end{flushright}
  \bigskip

  \begin{center}

                  \baselineskip=13pt {\LARGE \scshape{
                  Celestial Recursion
                  }
         }

      \vskip1cm

   \centerline{  {Yangrui Hu}${}^{1,2,3}$ and {Sabrina Pasterski}${}^{3,4}$ 

   }

\bigskip\bigskip
 \bigskip\bigskip
 
\centerline{\em${}^1$ {Department of Physics,
	Brown University,
	Providence, RI 02912, USA}
	}
\vspace{.5em}

 \centerline{\em${}^2$   Brown Theoretical Physics Center, Barus Hall, Providence, RI 02912,
USA
}
 \vspace{.5em}
 \centerline{\em${}^3$   
 Perimeter Institute for Theoretical Physics,
 Waterloo, ON N2L 2Y5, Canada
 }
\vspace{.5em}

 \centerline{\em${}^4$ Princeton Center for Theoretical Science, Princeton, NJ 08544, USA}

\bigskip\bigskip

\end{center}

\begin{abstract}

We examine the BCFW recursion relations for celestial amplitudes and how they inform the celestial bootstrap program. We start by recasting the celestial incarnation of the BCFW shift as a generalization of the action of familiar asymptotic symmetries on hard particles, before focusing on two limits: $z\to \infty$ and $z\to 0$.  We then discuss how the celestial CFT data encodes the large-$z$ behavior determining which shifts are allowed, while the infinitesimal limit is tied to the celestial bootstrap program via the BG equations that constrain the MHV sector.
The extension to super-BCFW is also presented.
We close by remarking on several open questions for future study.

\end{abstract}

 \bigskip \bigskip \bigskip \bigskip

\end{titlepage}

 \setcounter{tocdepth}{2}

\tableofcontents

\section{Introduction}

The Celestial Holography program~\cite{Pasterski:2021raf} aims to establish a duality between quantum gravity in asymptotically flat spacetimes and a codimension-2 celestial CFT (CCFT). It emerged from the observation that asymptotically flat spacetimes support infinite-dimensional symmetry enhancements that are made manifest in a scattering basis where the external particles are in boost eigenstates~\cite{Strominger:2017zoo}.  The objective of this program is to combine tools from 2D CFT and the 4D ${\cal S}$-matrix program: bootstrapping the 2D theory from its spectrum, symmetries, and OPE coefficients amounts to bootstrapping amplitudes from their soft and collinear limits.

Much of the recent activity in this vein has focused on the celestial symmetry algebras~\cite{Fan:2019emx,Fotopoulos:2019tpe,Fotopoulos:2019vac,Guevara:2021abz} that arise from collinear limits of scattering~\cite{Pate:2019lpp,Himwich:2021dau}, and how they constrain CCFT correlators~\cite{Banerjee:2020kaa,Banerjee:2020vnt,Hu:2021lrx,Fan:2022vbz}.    A common theme in these papers is that they are trying to rephrase features of scattering in a language familiar from 2D CFTs. However, the fact that the symmetry structure is much larger than what we would expect in an ordinary 2D CFT hints that there is value to taking a second look at the tools we have for momentum space amplitudes, to see how we might reorganize these enhancements.  For example,  when restricted to a single helicity sector the symmetry algebra for gravity is enhanced well beyond the asymptotic analyses of~\cite{Bondi:1962px,Sachs:1962wk,Sachs:1962zza,Barnich:2011ct,Barnich:2011mi} to include a $w_{1+\infty}$ symmetry~\cite{Strominger:2021mtt}.  This $w_{1+\infty}$ symmetry has recently been understood as arising from symmetries of self-dual gravity~\cite{Adamo:2021lrv}. It's Yang-Mills counterpart led to the {\it form factor integrand} construction of gauge amplitudes in~\cite{Costello:2022wso},  reminiscent of an unintegrated CSW~\cite{Cachazo:2004kj} formalism, in which the celestial chiral algebra plays a central role.

Both the on-shell recursion~\cite{Cachazo:2004kj,Britto:2004ap,Britto:2005fq,Arkani-Hamed:2008bsc,Nguyen:2009jk,Hodges:2011wm} and celestial CFT constructions are trying to identify a better way to compute $\mathcal{S}$-matrix elements than ordinary perturbation theory, where an unwieldy number of Feynman diagrams collapse to surprisingly simple answers.  While both routes start from little group covariance and the three-point data, the way that we build up higher point amplitudes from lower point amplitudes using the BCFW recursions is quite different from the approach one would take using the operator product expansion in CFT.  However, the main ingredients of the celestial bootstrap -- including soft limits, celestial OPEs, as well as the BG equations~\cite{Banerjee:2020zlg,Banerjee:2020vnt} which constrain MHV correlators -- can all be derived using BCFW shifts~\cite{Guevara:2019ypd,Pate:2019mfs,Pate:2019lpp,Himwich:2021dau,Hu:2021lrx}. The aim of this paper is to see what further insights we can gain into how to organize the celestial bootstrap program by revisiting these amplitude recursion relations.

The new results are as follows.  First, we recast the celestial incarnation of the BCFW recursion relations as an energy-dependent generalization of the superrotation asymptotic symmetry.
Second, we revisit the large-$z$ behavior and discuss how the celestial OPE data (coefficients and fusion rules) together with the little group scaling can tell us if the shifts are of an allowed type. Third, we show that these recursions imply an infinite tower of PDEs for tree-level MHV amplitudes that exponentiate the BG equations. Meanwhile, our asymptotic symmetry interpretation of the BCFW shift in terms of hard transformations and soft insertions yields the BG null state in the infinitesimal limit. This interpretation is straightforwardly extended to super-BCFW.

We begin with a review of the BCFW recursion relations~\cite{Britto:2005fq} in both the standard and celestial~\cite{Pasterski:2017ylz,Guevara:2019ypd} presentations in section~\ref{sec:BCFW}, emphasizing how the BCFW shifts generalize the kinds of deformations of the external kinematics associated with asymptotic symmetry transformations. We then discuss how these have been used to extract soft, collinear, and large-$z$ limits in section~\ref{sec:SClim}, and relate these to celestial bootstrap via BG equations in section~\ref{sec:BG}. Finally we close by discussing some open questions for future work in section~\ref{sec:disccussion}.
Our spinor helicity conventions and some helpful explicit computations are included in the appendix.

\section{From BCFW to Celestial BCFW}\label{sec:BCFW}

In this section, we review the standard momentum space construction for the BCFW recursion relations and show how it maps to the celestial basis. We will restrict to massless scattering in $(2,2)$ signature throughout. Before getting started, let us briefly establish our notation. We can parameterize the external momenta via
\be\label{eq:pi}
p^\mu_i=\epsilon_i\omega_i(1+z_i\bz_i,z_i+\bz_i,z_i-\bz_i,1-z_i\bz_i)
\ee
where $\epsilon_i=\pm 1$ indicates states that are outgoing or incoming when analytically continued from $\mathbb{R}^{1,3}$. In terms of spinor helicity variables this becomes
\be
p_i=-\lambda_i\tilde{\lambda}_i.
\ee
Our conventions are summarized in appendix~\ref{app:conventions}.  To go to the celestial basis we perform the following Mellin transform
\be
\label{eq:mellin}
\tilde{A}(\Delta_i,J_i,z_i,\bz_i)= \left(\prod_i \int_0^\infty d\omega_i\omega_i^{\Delta_i-1}\right)\,A(\omega_i,s_i,z_i,\bz_i).
\ee
This object behaves like a correlator of conformal primaries on the celestial sphere 
\be\label{eq:corr}
\tilde{A}(\Delta_i,J_i,z_i,\bz_i) =  {\cal N} \, \Bigl\langle \mathcal{O}^{\pm}_{\Delta_1,J_1}(z_1,\bz_1)\ldots \mathcal{O}^{\pm}_{\Delta_n,J_n}(z_n,\bz_n) \Bigr\rangle
\ee
with weights $(h_i,\bar{h}_i) = (\frac{\D_i+J_i}{2},\frac{\D_i-J_i}{2})$, where $J_i=s_i$ is the helicity and the normalization factor $  {\cal N} $ can be taken to be 1 for our purposes. We will focus on the gauge theory $(|s_i|=1)$ and gravity $(|s_i|=2)$ cases here.

\subsection{Momentum Space Recursion Relations}

The $\mathcal{S}$-matrix is a function of on-shell data, encoded in the spinors $\{\lambda_i,\tilde{\lambda}_i\}$ for massless scattering in 4D.  If we have control of the analytic structure of the amplitude, we can consider complex deformations of the external kinematics and apply tools from complex analysis to rewrite the amplitude in a useful way. We will be most interested in the BCFW shift, which acts on two spinors as follows
\begin{equation}
    \hat{\lambda}_i ~=~ \lambda_i ~+~ z\,\lambda_j ~~~,~~~ \hat{\tilde{\lambda}}_j ~=~ \tilde{\lambda}_j ~-~ z\,\tilde{\lambda}_i,~~~z\in\mathbb{C}
    \label{equ:bcfw-shift}
\end{equation}
and preserves the on-shell and momentum conservation constraints.  More general complex shifts of the external momenta 
\be\label{allshifts}
\hat{p}_i^\mu=p_i^\mu+zr_i^\mu,~~~\sum_i r_i^\mu=r_i\cdot r_j=p_i\cdot r_i=0,~~~z\in\mathbb{C}
\ee
can also be used. Here we will focus on tree-level amplitudes whose analytic structure is simpler.  Namely, we will only encounter simple poles and no branch cuts in $z$.

\begin{figure}[thb]
\centering
\vspace{.5em}
\begin{tikzpicture}[scale=2.1]
\draw[thick] (-1,0) --(1,0);
\draw[thick] (-2,.5) -- (-1,0);
\draw[thick] (-2,.3) -- (-1,0);
\draw[thick] (2,.5) -- (1,0);
\draw[thick] (-2,-.5) -- (-1,0);
\draw[thick] (2,-.5) -- (1,0);
\draw[thick] (2,-.3) -- (1,0);
\draw[fill] (-2,0)  circle (.02em);
\draw[fill] (-2,-0.1)  circle (.02em);
\draw[fill] (-2,-0.2)  circle (.02em);
\draw[fill] (2,0)  circle (.02em);
\draw[fill] (2,0.1)  circle (.02em);
\draw[fill] (2,0.2)  circle (.02em);
\draw[thick, fill=gray!50!white] (-1,0)  circle (1em);
\draw[thick, fill=gray!50!white] (1,0)  circle (1em);
\node[] at (-1,0) {L};
\node[] at (1,0) {R};
\node[] at (0,.25) {$\hat{P}_I$};
\end{tikzpicture}
\caption{ 
Complex shifts of the external kinematics can be used to write recursion relations for amplitudes. Here we show the contribution from the $z_I$ pole, where $\hat{P}_I^2=0$ and the shifted amplitude factorizes into two on-shell sub-amplitudes $\hat{A}_L$ and $\hat{A}_R$.
}
\label{fig:bcfw}
\end{figure}
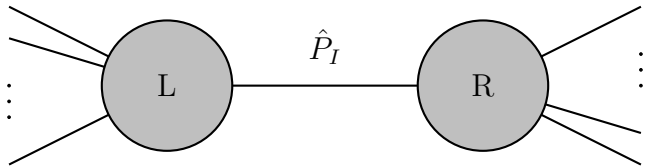

Let $A_n[1^{s_1}2^{s_2}3^{s_3}\cdots n^{s_n}]$  denote an $n$-point tree-level amplitude where $s_i$ labels the $i^{th}$ helicity.  The unshifted $n$-point amplitude can be written as the residue of the meromorphic function $z^{-1}\hat{A}_n(z)$ at the simple pole $z=0$
\begin{equation}
    \begin{split}\label{BCFWAn}
        A_n ~=~ \hat{A}_n(0) ~=&
        ~\oint_{|z|<\epsilon}\,\frac{dz}{2\pi i}\,\frac{\hat{A}_n(z)}{z} \\
        ~=&~ -\,\sum_{I}\,{\rm Res}_{z=z_I}\,\frac{\hat{A}_n(z)}{z} ~+~ B_n~.
    \end{split}
\end{equation}
To get the second line we've deformed the contour and used Cauchy's theorem. $B_n$ is the residue at $z=\infty$ which we will assume vanishes for the shifts under consideration in this section.\footnote{We will discuss the large-$z$ behavior in detail in section~\ref{sec:largez}.} Our task is to identify all of the poles $z_I$.  These can appear whenever an internal propagator goes on-shell, so that the sum in~\eqref{BCFWAn} is over all possible factorization channels. For our choice of shifts the deformed propagators are linear in $z$ 
\begin{equation}
    \hat{P}_I^2 ~=~ -\frac{P_I^2}{z_I}\,(z-z_I)~~
    \label{equ:hatP_I-onshell}
\end{equation}
and the residue of the shifted amplitude factorizes into a product of two lower-point amplitudes, as illustrated in figure~\ref{fig:bcfw}
\begin{equation}
    A_n ~=~ \sum_{I}\,\hat{A}_L(z_I)\,\frac{1}{P_I^2}\,\hat{A}_R(z_I)~.
    \label{equ:bcfw-momentum}
\end{equation}
We thus see that, when $B_n=0$, we get a recursion relation where we can build up higher-point amplitudes from lower-point amplitudes glued together at factorization channels that appear for complexified kinematics.

The factorization channels can be divided into two cases: collinear (for which $\hat{A}_L$ or $\hat{A}_R$ is a 3-point amplitude) and non-collinear, so that
$A_n=A_n^{\rm c}+A_n^{\rm nc}$. We will be most interested in the collinear sector for our purposes here.  Upon performing an $\langle 1,n]$ shift, the contribution from this sector takes the form
\begin{equation}
\begin{split}
    A_n^{\rm c} ~=&~ \sum_{i=2}^{n-1}\,\hat{A}_{3}[\hat{1}^{s_1},\,i^{s_i},\,-\hat{P}_{1i}^{-s_I}](z_i)\,\frac{1}{P_{1i}^2}\,\hat{A}_{n-1}[\hat{P}_{1i}^{s_I},\,\dots,\,\hat{n}^{s_n}](z_i)
\end{split}
\label{equ:collinear-generic-formula}
\end{equation}
where $z_i = \frac{\langle 1i \rangle}{\langle in \rangle}$. We can write this more explicitly using the spinor helicity variables. Recall that the 3-point amplitude is fixed by little group scaling arguments to take the form\footnote{Note that due to our choice of shift $\langle 1,n]$, if $s_1+s_2+s_3<0$ the shifted 3-point amplitude $\hat{A}_{3}$ vanishes. For the case $s_1+s_i-s_I<0$, we would instead choose a $[ 1,n\rangle$-shift.}
\begin{equation}\label{3ptsquare}
    A_3[1^{s_1}2^{s_2}3^{s_3}] ~=~ g_{123}\,[12]^{s_1+s_2-s_3}[32]^{s_2+s_3-s_1}[13]^{s_3+s_1-s_2}~~,~~~~~s_1+s_2+s_3>0~.
\end{equation}
We then have
\begin{equation}\scalemath{0.97}{
    \badat{3}
        A_n^{\rm c} ~=&~ \sum_{i=2}^{n-1}\,g_{1iI}\frac{[1i]^{s_i-s_I+s_1-1}\langle ni\rangle^{s_1-s_i+s_I}}{\langle n1\rangle^{s_1-s_i+s_I}\langle 1i \rangle}\,{A}_{n-1}\left[\lambda_i,\tilde{\lambda}_i+\frac{\langle n1 \rangle}{\langle ni \rangle}\tilde{\lambda}_1,s_I;\,\dots;\,\lambda_n,\tilde{\lambda}_n+\frac{\langle i1 \rangle}{\langle in \rangle}\tilde{\lambda}_1,s_n\right]
   \eadat}
    \label{equ:collinear-solved3pt}
\end{equation}
where we've used the convention $|-p\rangle = -|p\rangle$ and $|-p] = |p]$  for analytic continuation between channels. An important theme we will encounter throughout this paper is that this $A_n^c$ term gives us a lot of mileage: we can use it not only to extract soft, collinear, and (eventually) large-$z$ limits, but also to study the full MHV sector of gauge theory and gravity.

\subsection{Celestial Recursion Relations}\label{sec:celestial-recursion}

 The BCFW recursion relations played an important part in the derivation of the subleading soft graviton theorem~\cite{Cachazo:2014fwa} which, in turn, motivated the celestial basis. In this section we will revisit the BCFW shift with CCFT in mind.  We start by examining the role of angle-dependent Lorentz transformations in the BCFW story.  We will then review how to explicitly implement the BCFW shift in the celestial basis, following the presentation in~\cite{Guevara:2019ypd}.\footnote{For an early example restricted to the 4-point gluon amplitude see~\cite{Pasterski:2017ylz}.} 

\paragraph{From BCFW to CCFT}
Let us start with an $\langle i,j]$ BCFW shift. 
We can implement it with the following differential operator~\cite{Hu:2021lrx}
\begin{equation}\label{eq:dij}
    D_{i,j} ~=~ \lambda_j\,\frac{\pa}{\pa \lambda_i} ~-~ \Tilde{\lambda}_{i}\frac{\pa}{\pa \Tilde{\lambda}_{j}} 
\end{equation}
so that
\begin{equation}\label{eq:hatA}
    \exp\,\Big(z\, D_{ij}\Big) \, {A} ~=~ \hat{{A}}(z)
\end{equation}
shifts the spinors as in~\eqref{equ:bcfw-shift} and thus the amplitude, as desired.  As in~\cite{Guevara:2019ypd}, we can write this operator in terms of the angular momentum generator for particles $i$ and $j$. The global Lorentz generator
\be
J_{\mu\nu}\sigma^\mu_{\alpha\dot\alpha}\sigma^\nu_{\beta\dot\beta}=-2J_{\alpha\beta}\epsilon_{\dot\alpha\dot\beta}-2J_{\dot\alpha\dot\beta}\epsilon_{\alpha\beta}
\ee
splits into a chiral and an anti-chiral part
\be\label{jalphabeta}
J_{\alpha\beta}=\frac{i}{2}\left(\lambda_\alpha\frac{\p}{\p\lambda^\beta}+\lambda_\beta\frac{\p}{\p\lambda^\alpha}\right),~~~J_{\dot\alpha\dot\beta}=\frac{i}{2}\left(\tilde{\lambda}_{\dot \alpha}\frac{\p}{\p\tilde{\lambda}^{\dot\beta}}+\tilde{\lambda}_{\dot \beta}\frac{\p}{\p\tilde{\lambda}^{\dot\alpha}}\right)
\ee
which, when acting on a given scattering amplitude, can be further decomposed into generators $J^i$ acting on each external particle separately. A particular Lorentz transformation is specified by a choice of parameter $\theta_{\mu\nu}$
\be
U(\theta)=e^{i\theta_{\mu\nu}J^{\mu\nu}},~~~~\theta_{\mu\nu}\sigma^\mu_{\alpha\dot\alpha}\sigma^\nu_{\beta\dot\beta}=-2\theta_{\alpha\beta}\epsilon_{\dot\alpha\dot\beta}-2\theta_{\dot\alpha\dot\beta}\epsilon_{\alpha\beta}~.
\ee
We will also consider the direction-dependent generalization 
\be\label{gen_U}
U(\vec{\theta})=\exp\Big[\,\sum_k\,\theta_k^{\a\b}\,\left(\lambda_{k,\alpha}\frac{\p}{\p\lambda_k^\beta}+\lambda_{k,\beta}\frac{\p}{\p\lambda_k^\alpha}\right)+\theta_k^{\dot\a\dot\b}\,\left(\tilde{\lambda}_{k,\dot \alpha}\frac{\p}{\p\tilde{\lambda}_k^{\dot\beta}}+\tilde{\lambda}_{k,\dot \beta}\frac{\p}{\p\tilde{\lambda}_k^{\dot\alpha}}\right)\,\Big]~.
\ee
For a global Lorentz transformation we need $\theta^{\mu\nu}_k=\theta^{\mu\nu}$ to be the same for all of the external particles. If, instead, we were free to make the choice
\be\label{eq:theta}
\theta_i^{\alpha\beta}=z\frac{|j\rangle^{\alpha}|j\rangle^{\beta}}{2\,\langle ij\rangle}~~,~~~\theta_j^{\dot\alpha\dot\beta}=z\frac{[i|^{\dot\alpha}[i|^{\dot\beta}}{2\,[ij]}~~,
\ee
with all other $k\neq\{i,j\}$ vanishing, the operator~\eqref{gen_U} would indeed implement~\eqref{eq:dij}. However, once we restrict to only some subset of the external particles and complexify the generators we are going beyond ordinary Lorentz transformations. 
Let's persist nonetheless. Noting that \be
\epsilon^\mu_{-,j}\sigma_\mu=\frac{|r]\langle j|}{[ j r]},~~~
\epsilon^\mu_{+,j}\sigma_\mu=
\frac{|j]\langle r|}{\langle j r\rangle}
\ee 
we can write these shifts in a more recognizable form (no sum over $i,j$)
\be
\theta_i^{\mu\nu}J^i_{\mu\nu}= -2
z\langle ij\rangle^{-1}{\epsilon_{-,j}^\mu p_j^\nu J^i_{\mu\nu}},~~~\theta_j^{\mu\nu}J^j_{\mu\nu}=2z[ij]^{-1}{\epsilon_{+,i}^\mu p_i^\nu J^j_{\mu\nu}}
\ee
where we note that the reference spinor dependence drops out.  By restricting to the chiral or anti-chiral part of~\eqref{eq:dij}, one can generalize this to an arbitrary shift in the spinor variables
\be\label{eq:lorentzparam}
\lambda_i\mapsto \lambda_i+z\lambda_s ~~\Leftrightarrow ~~\theta^{\mu\nu}_j= -2z\delta_{ij}\langle js\rangle^{-1}\epsilon_{-,s}^\mu p_s^\nu~,
\ee
and similarly for $\tilde{\lambda}_i$ with the polarization choice flipped.  This operator exponentiates because it has been designed to act trivially on all but the $i^{th}$ spinor.  

Let us now try to interpret this notion of modulating our Lorentz transformations as a function of the celestial sphere. We see that for the BCFW recursion relations we want to be able to perform a different Lorentz transformation on each particle.  However, in quantum field theory, we only have the Poincar\'e isometries.  Adding gravity changes this.  The asymptotic symmetry group of asymptotically flat spacetimes includes superrotations\footnote{To connect to the asymptotic symmetry story, we will switch to $(1,3)$ signature in this section, for which the complex conjugate terms are the corresponding anti-chiral transformations. Since we want to complexify the symmetry action in the end, we can just as well drop the + c.c. terms in this context.}
\be\badat{3}\label{xiy}
\xi&=(1+\frac{u}{2r})Y^z\p_z-\frac{u}{2r}D^\bz D_z Y^z\p_\bz-\frac{1}{2}(u+r)D_zY^z\p_r +\frac{u}{2}D_zY^z\p_u+c.c.
\eadat
\ee
which extend the global Lorentz transformations
\be\begin{array}{lll}\label{lorentz}
   Y^z_{12}=iz,~~  & Y^z_{13}=-\frac{1}{2}(1+z^2),~~ & Y^z_{23}=-\frac{i}{2}(1-z^2) \\
   Y^z_{03}=z,~~  & Y^z_{02}=-\frac{i}{2}(1+z^2),~~ & Y^z_{01}=-\frac{1}{2}(1-z^2) \\
\end{array}
\ee
to local conformal transformations, allowing meromorphic $Y^z=Y(z)$. The subleading soft graviton theorem~\cite{Cachazo:2014fwa} implies that the perturbative gravitational $\cal S$-matrix obeys a Ward identity~\cite{Kapec:2014opa}
\be\label{wardid}
\langle out | Q^+[\xi] \mathcal{S}-\mathcal{S}Q^-[\xi]|in\rangle=0,~~~Q^\pm[\xi]=Q_S^\pm[\xi]+Q_H^\pm[\xi]
\ee
for superrotations that acts diagonally on the in and out states. Here the soft charge inserts a subleading soft graviton 
\be\label{Qs}
Q_S^+[\xi]\propto \lim\limits_{\omega\rightarrow0}(1+\omega\p_\omega)\int d^2 z\sqrt{\gamma} D_z^3Y^z a_{-}(\omega)+h.c.
\ee
while the hard charge $Q^\pm_H[\xi]=i\mathcal{L}_{\xi}$ implements the corresponding coordinate transformation on the hard particles
\be\label{QH}
\mathcal{L}_{\xi}|\omega_k,z_k,\bz_k\rangle= Y^z_k\p_{z_k}-\frac{1}{2}D_zY^z(-\omega_k\p_{\omega_k}+s_k)+h.c.
\ee
One can show that the subleading soft factor reduces to the form~\eqref{QH} when we integrate~\eqref{Qs} by parts. The necessary relations to go from the spinor representation of the angular momentum~\eqref{jalphabeta} to the energy-momentum presentation used here can be found in appendix~\ref{app:conventions}.    

Now let us return to the BCFW shift that motivated this interlude.  In the case where we are dealing with a finite number of external scatters, we can try to approximate any specified Lorentz transformation in the vicinity of where each of the $n$ massless particles punctures the celestial sphere with a particular choice of meromorphic $Y(z)$. For example, we can construct a $Y$ that behaves like
\be\label{eq:de}
Y(z)|_{z\sim z_i}=\alpha_i+\beta_i(z-z_i)+\gamma_i(z-z_i)^2+...
\ee
near each $i\in\{1,...,n\}$, matching the local form of a Lorentz transformation~\cite{srpi}. 
The $\alpha_i,\beta_i,\gamma_i\in\mathbb{C}$ would determine the $\theta^i_{\mu\nu}$ specifying an element of SL(2,$\mathbb{C}$). This freedom to independently boost and rotate well-separated jet directions is closely related to reparameterization invariance in Soft Collinear Effective Theory (SCET)~\cite{Bauer:2000yr,Bauer:2001ct,Bauer:2001yt}. As we will explore in more detail in appendix~\ref{app:ASG}, if we compare our BCFW transformation parameters~\eqref{eq:theta} to~\eqref{QH} we have
\be
\theta_i^{\alpha\beta}=\frac{|j\rangle^{\alpha}|j\rangle^{\beta}}{2\,\langle ij\rangle}~~\Leftrightarrow~~ Y|_{z\sim z_i}\propto \sqrt{\frac{\omega_j}{\omega_i}}(z_{ij}+2(z-z_i)+...)
\ee
near particle $i$, and vanishing near the other particle positions.  While the transformation of the left-handed spinors is meromorphic in the $z_k$ and we can match the local form of the transformation to a complexified superrotation, the energy dependence of the parameters indicates this is clearly a generalization of what is familiar from the spacetime asymptotic symmetry group.\footnote{If instead we wanted our transformation to match the subleading soft factor we would see that the amount by which we want to boost each particle is a function of the angles not the energies of the particles. However, unlike for the $\langle i,j]$ shift, the corresponding $Y^z$ is non-meromorphic as a function of the direction of the soft graviton.  Instead, the subleading soft theorem is isomorphic to the Ward identity of a Diff$(S^2)$ transformation related by a shadow transform to this stress tensor~\cite{Donnay:2020guq,Pasterski:2021fjn}.} 
Because the celestial basis trades ordinary energies $\omega_k$ for Rindler energies $\omega_k\p_{\omega_k}$, we will need to treat these energy-dependent prefactors as weight-shifting operators.  Alternatively, to have energy-independent parameters, we can consider more general shifts of the external kinematics which is easier to implement if we relax the momentum conservation constraint (i.e. only require the on-shell condition is preserved). This interpretation of the BCFW shift as a generalization of the action of superrotations on hard particles also suggests a relation to soft insertions, which we will explore in detail in section~\ref{sec:BG}.

\paragraph{Celestial BCFW}
Above we contrasted our ability to independently shift the on-shell spinors for particles traveling in different directions to the  structure of the asymptotic symmetry group of asymptotically flat spacetimes.  The fact that these symmetries resemble those of a 2D CFT is what motivated the celestial holography program.  While Lorentz invariance guarantees that we can look at scattering in a boost basis, the non-trivial content of the soft graviton theorem is that the quasi-primary external operators get promoted to Virasoro primaries when we turn on gravity in the bulk. Now we will turn our attention to constructing the BCFW recursion formula in this basis by first rewriting the BCFW shift (\ref{equ:bcfw-shift}) as shifts in ($\epsilon$, $\o$, $z$, $\bz$) following \cite{Guevara:2019ypd}. Since we are interested in extracting the CCFT data and differential equations constraining the MHV sector, we will focus on formulating the collinear channel recursion in what follows.  

Recall that the Lorentz group in 4D acts on our left- and right-handed spinors as follows
\begin{equation}
    |{\lambda}'\rangle ~=~ \Lambda\,|\lambda\rangle ~~~,~~~ |{\lambda}'] ~=~ \tilde{\Lambda}\,|\lambda]
\end{equation}
where
\begin{equation}
    \Lambda ~=~ \begin{pmatrix}
    d & c\\
    b & a\\
    \end{pmatrix} ~~~,~~~
    \tilde{\Lambda} ~=~ \begin{pmatrix}
    \bar{a} & -\bar{b}\\
    -\bar{c} & \bar{d}\\
    \end{pmatrix} ~~~,~~~ 
    ad-bc ~=~ \bar{a}\bar{d}-\bar{b}\bar{c} ~=~ 1~.
    \label{equ:Lorentztransfmatrix}
\end{equation}
In $(1,3)$ signature $\bar a=a^*$ and so forth, while in $(2,2)$  signature $\Lambda,\tilde{\Lambda}$ are two independent SL(2,$\mathbb{R})$-valued matrices. Using (\ref{equ:spinorhelicity-celestial}), we have
\begin{equation}
    \begin{split}
     |{\lambda}'\rangle~=&~  \Lambda\,\epsilon\sqrt{2\o t}\begin{pmatrix}
        1\\
        z
        \end{pmatrix} ~=~ \sqrt{\left|\frac{cz+d}{\bar{c}\bar{z}+\bar{d}}\right|}\,\epsilon'\,\sqrt{2\o' t}\begin{pmatrix}
        1\\
        z'
        \end{pmatrix}\\
        |{\lambda}'] ~=&~ \tilde{\Lambda}\,\sqrt{2\o t^{-1}}\begin{pmatrix}
        -\bar{z}\\
        1
        \end{pmatrix} ~=~ \sqrt{ \left|\frac{\bar{c}\bar{z}+\bar{d}}{cz+d}\right|}\,\sqrt{2\o' t^{-1}}\begin{pmatrix}
        -\bar{z}'\\
        1
        \end{pmatrix}
    \end{split}
    \label{equ:lorentz-little-group}
\end{equation}
where
\begin{equation}
    z'~=~ \frac{az+b}{cz+d}~~,~~\bar{z}'~=~ \frac{\bar{a}\bar{z}+\bar{b}}{\bar{c}\bar{z}+\bar{d}}\label{mobius}
\end{equation}
and
\begin{equation}
    \o' ~=~ \o\,|cz+d||\bar{c}\bar{z}+\bar{d}| ~~,~~ \epsilon' ~=~ \epsilon\,{\rm sgn}(cz+d){\rm sgn}(\bar{c}\bar{z}+\bar{d}). \label{omegaeps}
\end{equation}
Here $t$ is a little group parameter, which is commonly gauge fixed to 1. We further note that the reality conditions in $(1,3)$ signature imply that $\epsilon$ is a Lorentz invariant, however this is not true in $(2,2)$.  

Little group scaling dictates that if we transform the spinor for a particle with spin $s$, the ordinary momentum space amplitude transforms as follows
\begin{equation}
    A(|{\lambda}'\rangle,\,|{\lambda}'],\,s) ~=~ \left(\sqrt{\left|\frac{cz+d}{\bar{c}\bar{z}+\bar{d}}\right|}\right)^{-2s}\,A(\epsilon',\o',z',\bar{z}')~.
\end{equation}
The Mellin transform~\eqref{eq:mellin} has the effect of further diagonalizing the boost weight 
\begin{equation}\scalemath{1}{
    \Lambda_i\,\tilde{\Lambda}_i\,\tilde{A}(\epsilon_i,\D_i,J_i,z_i,\bar{z}_i) ~=~ |c z_i+d|^{-2h_i}|\bar{c}\bar{z}_i+\bar{d}|^{-2\bar{h}_i}\,\tilde{A}\left(\epsilon_i',\D_i,J_i,z_i',\bz_i'\right).}
    \label{equ:conf-covariance}
\end{equation}
Note that here we are considering the transformation acting on only one particle in contrast to doing a global Lorentz transformation. 
From~\eqref{eq:mellin}, we also see that in this basis the energies get promoted to operators that shift the corresponding conformal dimensions
\be
 \omega_i \times (\cdot) ~~\mapsto~~T_i(\cdot),~~~{\rm where} ~~~T_i: \Delta_i\mapsto \Delta_i+1.
\ee
With this it is straightforward to translate the operator $D_{i,j}$ defined in~\eqref{eq:dij} to the celestial basis. We have
\begin{equation}
    D_{i,j} ~=~-\,\Omega_{i,j}\,\left(\,2h_i~+~z_{ij}\frac{\pa}{\pa z_i} \,\right) ~+~ \Tilde{\Omega}_{i,j}\,\left(\,2\Bar{h}_j~+~\Bar{z}_{ji}\frac{\pa}{\pa \Bar{z}_j} \,\right)
    \label{equ:Dij-CS}
\end{equation}
where
\begin{equation}\label{LambdatLambda}
    \Omega_{i,j} ~=~ \frac{\epsilon_j\,\sqrt{t_j}}{\epsilon_i\,\sqrt{t_i}}\,T_i^{-\frac{1}{2}}\,T_j^{\frac{1}{2}}~~~,~~~\Tilde{\Omega}_{i,j}~=~\frac{\sqrt{t_j}}{\sqrt{t_i}}\,T_i^{\frac{1}{2}}\,T_j^{-\frac{1}{2}}.
\end{equation}
Because the BCFW shift preserves the momentum conservation locus, this operator can act on either the momentum-conservation stripped or the full on-shell amplitude.  

Equation~\eqref{eq:hatA} then implements the corresponding $z$-dependent shifted amplitude $\hat{A}(z)$ to which we can apply the residue analysis in~\eqref{BCFWAn} to repackage $A_n$ in terms of factorization channels.  However, we can also apply the same procedure directly to the result, since in that case we are just applying the shifts for specific $z=z_I$. Since the collinear contribution~\eqref{equ:collinear-solved3pt} will be most interesting in what follows, let us translate that expression directly to the celestial basis.

Starting with (\ref{equ:collinear-solved3pt}), the Lorentz transformations acting on the $A_{n-1}$ are
\begin{align}
\label{equ:Lorentztransfi}
 {\Lambda}_i ~=&~ \mathbb{1}_2 ~~,~~  \tilde{\Lambda}_i ~=~ \mathbb{1}_2 + \frac{\langle n1 \rangle}{\langle ni \rangle[1i]}|1][1|
  ~=~
  \begin{pmatrix}
1-\frac{\a}{\bar{z}_{1i}}\bar{z}_1 & -\frac{\a}{\bar{z}_{1i}}\bar{z}^2_1 \\
  \frac{\a}{\bar{z}_{1i}} & 1+\frac{\a}{\bar{z}_{1i}}\bar{z}_1
    \end{pmatrix}\\
  {\Lambda}_n ~=&~ \mathbb{1}_2 ~~,~~  \tilde{\Lambda}_n ~=~ \mathbb{1}_2 + \frac{\langle i1 \rangle}{\langle in \rangle[1n]}|1][1|
  ~=~
  \begin{pmatrix}
  1-\frac{\b}{\bar{z}_{1n}}\bar{z}_1 & -\frac{\b}{\bar{z}_{1n}}\bar{z}^2_1 \\
  \frac{\b}{\bar{z}_{1n}} & 1+\frac{\b}{\bar{z}_{1n}}\bar{z}_1
    \end{pmatrix}\label{equ:Lorentztransfn}
\end{align}
where the parameters $\alpha$ and $\beta$ are related to the celestial coordinates as follows 
\begin{equation}
    \a ~=~ \frac{\epsilon_1 z_{n1}}{\epsilon_i z_{ni}}\,\o_1\,T_i^{-1} ~~~,~~~ 
    \b ~=~ \frac{\epsilon_1 z_{i1}}{\epsilon_n z_{in}}\,\o_1 \,T_n^{-1} ~.~~
\end{equation}
and we have written $\o_1$ in place of $T_1$ to emphasize that from the point of view of the $n-1$ point amplitude $\omega_1$ is a parameter and not an operator.
Comparing (\ref{equ:Lorentztransfi}) and (\ref{equ:Lorentztransfn}) with (\ref{equ:Lorentztransfmatrix}), we can identify the corresponding M\"obius transformation~\eqref{mobius} of the celestial coordinates
\begin{equation}
    \begin{split}
     \bar{z}'_i ~=~\bar{z}_i+\frac{\a\bar{z}_{1i}}{1+\a}~, ~~~ \bar{z}'_n ~=~\bar{z}_n+\frac{\b\bar{z}_{1n}}{1+\b}~. \\
    \end{split}
\end{equation}
Mapping (\ref{equ:collinear-solved3pt}) to the celestial sphere then directly gives us a celestial recursion relation
\begin{equation}
\begin{split}
        \tilde{A}_n^c ~=~ \int_0^{\infty}\frac{d\o_1}{\o_1}\,\o_1^{\D_1}\,\sum_{i=2}^{n-1}&\,g_{1iI}\,\frac{C_i(s_1)}{\o_1}\,|1+\a|^{-2\bar{h}_i}\,|1+\b|^{-2\bar{h}_n}\\
        &\tilde{A}_{n-1}\left(\epsilon_i',z_i,\bar{z}_i+\frac{\a\bar{z}_{1i}}{1+\a};\dots;\epsilon_n',z_n,\bar{z}_n+\frac{\b\bar{z}_{1n}}{1+\b} \right)
 \end{split}
    \label{equ:c-recursion}
\end{equation}
where $C_i(s_1)$ is a prefactor depends on the helicity of the first particle, 
\begin{align}\label{eq:Cs}
C_i(s_1) ~=&~ \frac{z_{ni}}{-2z_{1i}z_{n1}t_1}\,\left(2\,\frac{\epsilon_i}{\epsilon_1}\,\frac{z_{ni}\bar{z}_{1i}}{z_{n1}t_1}\,T_i \right)^{s_1-1}\,\left( 2\,\frac{\epsilon_1}{\epsilon_i}\,\frac{z_{n1}\bar{z}_{1i}}{z_{ni}t_i}\,\o_1\right)^{s_i-s_I} ~.
\end{align}
Alternatively, we can follow~\cite{Guevara:2019ypd} and embrace an operator implementation~\eqref{eq:hatA} of the respective shifts.  Using (\ref{equ:spinorhelicity-celestial})-(\ref{equ:map}), equation~\eqref{equ:c-recursion} then becomes
\begin{equation}
    \begin{split}
        \Tilde{A}_n^{\rm c} ~=&~ \int_0^{\infty}\frac{d\o_1}{\o_1}\,\o_1^{\D_1}\, \sum_{i=2}^{n-1}\,g_{1iI}\,\frac{C_i(s_1)}{\o_1}\\
        &\qquad \exp\left\{\,\o_1\,\frac{\epsilon_1 z_{n1}}{\epsilon_i z_{ni}}\,\Big[\,T_i^{-1}\,(\bar{z}_{1i}\bar{\pa}_i-2\bar{h}_i) +  \frac{\epsilon_i z_{1i}}{\epsilon_n z_{n1}}\,T_n^{-1}\,(\bar{z}_{1n}\bar{\pa}_n-2\bar{h}_n)\,\Big] \,\right\}\,\Tilde{A}_{n-1}
    \end{split}
    \label{equ:c-recursion-2}
\end{equation}
In practice, we can easily derive higher-point celestial MHV amplitudes from the 3-point function using \eqref{equ:c-recursion}. We will give an explicit example in appendix \ref{appen:4gluon}. Meanwhile, \eqref{equ:c-recursion-2} lends itself to extracting PDEs constraining celestial correlators, as we will see in section \ref{sec:mhvpde}.

\section{\texorpdfstring{Extracting Soft, Collinear, and Large-$z$ Limits}{Extracting Soft, Collinear, and Large-z Limits}}\label{sec:SClim} We start this section by reviewing how to use the output of the BCFW recursion relation, in the form~\eqref{equ:c-recursion-2}, to extract the soft~\cite{Guevara:2019ypd} and collinear~\cite{Pate:2019lpp} limits of the celestial amplitude.  These will provide us with the symmetries and OPE coefficients of the celestial CFT.  
We then show how the complexified collinear limits signal which shifts are allowed in section~\ref{sec:largez}. 
Here we will simply focus on extracting the relevant data. We will reassess how to interpret the BCFW manipulations from the point of view of the celestial bootstrap in the following section.

\subsection{Currents from Soft Limits}\label{sec:soft_currents}

Let us start with a generic amplitude and consider what happens when we take the external leg $s$ to be soft. Only the collinear factorization channels will contribute up to subleading order in gauge theory and sub-subleading order in gravity~\cite{Cachazo:2014fwa}. Without loss of generality, we can take $s_s>0$ and perform the BCFW shift 
\begin{equation}
    \hat{\lambda}_s ~=~ \lambda_s ~+~ z\,\lambda_n ~~~,~~~ \hat{\tilde{\lambda}}_n ~=~ \tilde{\lambda}_n ~-~ z\,\tilde{\lambda}_s~.
    \label{equ:bcfw-shift-soft}
\end{equation}
With a slight abuse of notation let $\tilde{A}^c_n(\omega_s)$ denote the integrand of~\eqref{equ:c-recursion-2}. We have
\begin{equation}
   \scalemath{0.94}{
    \badat{3}
        \Tilde{A}_n^{\rm c}(\omega_s) ~=
        \sum_{i=2}^{n-1}\,g_{isI}\,\frac{C_i(s_s)}{\o_s}\,\exp\left\{\,\o_s\,\frac{\epsilon_s z_{ns}}{\epsilon_i z_{ni}}\,\Big[\,T_i^{-1}\,(\bar{z}_{si}\bar{\pa}_i-2\bar{h}_i) +  \frac{\epsilon_i z_{si}}{\epsilon_n z_{ns}}\,T_n^{-1}\,(\bar{z}_{sn}\bar{\pa}_n-2\bar{h}_n)\,\Big] \,\right\}\,\Tilde{A}_{n-1}
 \eadat}
    \label{equ:mom-soft}
\end{equation}
where only $s_i = s_I$ will contribute.  The prefactors $C_i$ from~\eqref{eq:Cs} take the form
\begin{align}
 \text{soft gluon:} \qquad  C_i(s_s=1) ~=&~ \frac{z_{ni}}{-2z_{si}z_{ns}t_s} ~=~ \frac{1}{-2t_s}\left(\frac{1}{z_{si}}-\frac{1}{z_{sn}}\right)~~, \\
  \text{soft graviton:} \qquad   C_i(s_s=2) ~=&~ \frac{z_{ni}^2\bar{z}_{si}}{z_{is}z^2_{ns}t^2_s}\,\frac{\epsilon_i}{\epsilon_s}\,T_i~.
\end{align}
The helicity assignment  $s_i = s_I$  is consistent with the expectation that the soft theorems give a relation between amplitudes with and without an extra particle $s$, with the helicities of hard particles fixed.

Now let's consider what happens when we do the $\omega_s$ integral. The coefficients of the Laurent expansion near $\omega_s\rightarrow0$ translate into poles in the celestial amplitude.  In particular, the ${\o}^{-1}$, $\o^0$, and $\o$ orders of the $A_n^c$ expansion give poles at $\D_s=1,0,-1$ respectively~\cite{Cheung:2016iub,Pate:2019mfs}. This encompasses the soft theorems considered in~\cite{Cachazo:2014fwa}.   However, since the $A_n^c$ term is the only contribution for MHV amplitudes, we can continue our expansion to all orders in $\o_s$ for this case. This gives a tower of poles at $\D_s=1-\mathbb{Z}_+$~\cite{Guevara:2019ypd}. Let us define the conformal soft current $H^k$ as the residue 
\begin{equation}
    H^k\,\tilde{A}_{n-1} ~:= \lim\limits_{\D_s\to 1-k}\,(\D_s-1+k)\,\tilde{A}_n ~=~{\rm Res}_{\o=0}\,\o^{1-k}\,A_n~.
\end{equation}
Upon Taylor expanding the exponential in~\eqref{equ:mom-soft} and calculating the residues one finds the following operator expression for its action on MHV amplitudes
\begin{equation}
    \begin{split}
         H^k(J_s) ~=&~ \frac{1}{k!}\, \sum_{i=2}^{n-1}\,g_{isI}\,C_i(J_s)\,\left[\,\frac{\epsilon_s z_{ns}}{\epsilon_i z_{ni}}\,\Big(\,T_i^{-1}\,(\bar{z}_{si}\bar{\pa}_i-2\bar{h}_i) +  \frac{\epsilon_i z_{si}}{\epsilon_n z_{ns}}\,T_n^{-1}\,(\bar{z}_{sn}\bar{\pa}_n-2\bar{h}_n)\,\Big)\,\right]^{k}~.
    \end{split}
    \label{equ:soft-currents}
\end{equation}
We see that conformally soft limits of scattering can be recast as differential operators acting on the hard part of the amplitude. 

Before moving on, it is worth looking at an example to compare our notation to what we expect from the early celestial literature~\cite{Strominger:2017zoo}.  The leading soft gluon theorem reduces to
\begin{equation}\label{leadinggluon}
    \lim\limits_{\D\to 1}\,(\D-1)\,\langle {\cal O}_{\D,+}{\cal O}_{\D_2,J_2}\cdots{\cal O}_{\D_n,J_n} \rangle ~=~ \frac{g}{t_1}\,\frac{z_{n2}}{-2z_{12}z_{n1}}\,\langle {\cal O}_{\D_2,J_2}\cdots{\cal O}_{\D_n,J_n} \rangle~~,
    \end{equation}
 for the color stripped amplitudes.   We can restore the color factors for the full correlator
\be
  \langle\,{\cal O}^{a_1}\cdots{\cal O}^{a_n}\,\rangle ~=~ \sum_{\sigma \in S_{n-1}}\, \langle\, {\cal O}_{1}{\cal O}_{\sigma_2}\cdots{\cal O}_{\sigma_n}\,\rangle\,{\rm Tr}[{\cal T}^{a_1}{\cal T}^{a_{\sigma_2}}\cdots{\cal T}^{a_{\sigma_n}}]
\ee
where $\cal{T}^a$ denote a generator of the gauge group and $\mathcal{O}^b$ an operator in the adjoint representation 
\begin{equation}\label{Tadj}
    {\cal T}_i^a \,{\cal O}_j^b ~=~ i\,f^{abc}\,{\cal O}^c_j\,\d_{ij}~.
\end{equation}
 Taking $\mathcal{O}^{a_1}$ conformally soft and applying~\eqref{leadinggluon} gives
\begin{equation}\scalemath{0.95}{
   \badat{3}
        \langle\, H^{a,(0)}(z_s,\bz_s)\,{\cal O}^{a_2}\cdots{\cal O}^{a_n}\,\rangle ~=&~ \sum_{\sigma \in S_{n-1}}\,\left(\frac{1}{z_{s\sigma_2}}- \frac{1}{z_{s\sigma_n}}\right)\langle\, {\cal O}_{\sigma_2}{\cal O}_{\sigma_3}\cdots{\cal O}_{\sigma_n}\,\rangle\,{\rm Tr}[{\cal T}^{a}{\cal T}^{a_{\sigma_2}}{\cal T}^{a_{\sigma_3}}\cdots{\cal T}^{a_{\sigma_n}}]\\
        ~=&~ \sum_{i=2}^n\,\frac{1}{z_{si}}\,\sum_{\sigma\in S_{n-2}}\langle\, {\cal O}_{i}{\cal O}_{\sigma_3}\cdots{\cal O}_{\sigma_n}\,\rangle\,{\rm Tr}(\,[{\cal T}^{a},\,{\cal T}^{a_i}]\,{\cal T}^{a_{\sigma_3}}\cdots{\cal T}^{a_{\sigma_n}})\\
        ~=&~ \sum_{i=2}^n\,\frac{{\cal T}_i^a}{z_{si}}\,\langle\,{\cal O}^{a_2}\cdots{\cal O}^{a_n}\,\rangle
   \eadat}
\end{equation}
where we've used~\eqref{Tadj} and units where $g_{\rm YM}=1$. We see that the $k=0$ current generates a Kac-Moody symmetry on the celestial sphere~\cite{He:2015zea}. Meanwhile for $k=1$  we have
\begin{equation}
    \langle\, H^{a,(1)}(z_s,\bz_s)\,{\cal O}^{a_2}\cdots{\cal O}^{a_n}\,\rangle ~=~ -\, \sum_{i=2}^n\,\frac{\epsilon_s}{\epsilon_i z_{si}}\,\Big(\bz_{si}\bar{\pa}_i - 2\bar{h}_i +1 \Big)\,T_i^{-1}\,{\cal T}_i^a\,\langle\,{\cal O}^{a_2}\cdots{\cal O}^{a_n}\,\rangle
\end{equation}
which corresponds to the Yang-Mills analog of Low's subleading soft theorem~\cite{Low:1954kd,Low:1958sn}, whose celestial manifestation was studied in~\cite{Himwich:2019dug}. We can similarly extend these color factor-restored expressions to the full $k\in\mathbb{Z}_{\ge0}$ tower. 

\subsection{OPEs from Collinear Limits}\label{sec:collinear}

Because celestial amplitudes involve fixed angle scattering, the collinear limit in momentum space maps directly to the coincident limit on the celestial sphere. As such, knowledge of the collinear splitting functions (i.e. the contribution from $A^c$) is sufficient to determine the celestial OPE coefficients for massless external states~\cite{Pate:2019lpp,Fotopoulos:2019vac}.\footnote{This is implicit in the discussions of collinear limits in~\cite{Himwich:2021dau}, where they consider a shift that involves particles 1 and $k\neq 1,2$ while examining the collinear limit of particles 1 and 2. They then determine where a pole in the holomorphic collinear variable $z_{12}$ can appear.  At tree level, we know such a pole can arise from propagators in Feynman diagrams where a virtual particle splits into the collinear pair via a three-point interaction~\cite{Taylor:2017sph}. Here we are interested in the deformed amplitudes, so comparing to~\eqref{BCFWAn} we can also get a pole coming from the factor of $z^{-1}$. The only pole for which $z\propto z_{12}$ will come from a deformation such that $\langle\hat{1}2\rangle\rightarrow 0$, which is the collinear factorization channel.} In this section, we will show how to extract the celestial OPE coefficients corresponding to the leading holomorphic collinear singularity (scaling like $z_{12}^{-1}$) from a $\langle 1,n]$-shift.\footnote{The anti-holomorphic collinear singularity $\zb_{12}^{-1}$ can be extracted from a $[1,n\rangle$-shift}

For simplicity, we will focus on the case where $\epsilon_1 = \epsilon_2 = \epsilon_n = 1$. The $i=2$ term in (\ref{equ:collinear-generic-formula}) is the only one that can contribute to this leading OPE.  Using (\ref{equ:spinorhelicity-celestial}) we have
\begin{equation}
P_{12}^2 ~=~ (-4)\o_1\o_2z_{12}\bar{z}_{12} ~~~,~~~    \hat{\lambda}_1 ~=~ \sqrt{\frac{\o_1 t_1}{\o_2 t_2}}\,\frac{z_{1n}}{z_{2n}}\,\lambda_2~,
\end{equation}
while we can write $\hat{P}_{12} = -\lambda_P\tilde{\lambda}_P $with
\begin{equation}\label{eq:p12}
    \begin{split}
    \begin{array}{ll}
    \lambda_P ~=&~ \sqrt{\frac{\o_1 t_1+\o_2 t_2}{\o_2 t_2}}\lambda_2 \\ &\\
   \tilde{\lambda}_P  ~=&~  \sqrt{\frac{\o_1 t_1}{\o_1 t_1+\o_2 t_2}}\frac{z_{1n}}{z_{2n}}\tilde{\lambda}_1+ \sqrt{\frac{\o_2 t_2}{\o_1 t_1+\o_2 t_2}}\tilde{\lambda}_2 
    \end{array}
    ~~\leftrightarrow~~ \begin{cases}
    \o_P ~=~ \o_1\frac{z_{1n}}{z_{2n}}+\o_2\\
    t_P ~=~ \frac{\o_1 t_1+\o_2 t_2}{ \o_1\frac{z_{1n}}{z_{2n}}+\o_2}\\
    z_P ~=~ z_2 \\
    \bz_P ~=~ \frac{\o_1\,\frac{z_{1n}}{z_{2n}}\bar{z}_1+\o_2\bar{z}_2}{\o_1\,\frac{z_{1n}}{z_{2n}}+\o_2}~.
    \end{cases}
    \end{split}
\end{equation}
The 3-point shifted amplitude then takes the form
\begin{equation}
    \begin{split}
        \hat{A}_{3}[\hat{1}^{s_1},\,2^{s_2},\,-\hat{P}_{12}^{-s_P}](z_I) ~=&~ g_{12P}(2\bar{z}_{12})^{s_1+s_2-s_P}\,\o_1^{s_2-s_P}\,\o_2^{s_1-s_P}\\
        &~(\o_1 t_1+\o_2 t_2)^{s_P}\,t_1^{-s_1}\,t_2^{-s_2}\,\left(\frac{z_{1n}}{z_{2n}}\right)^{s_2-s_P-s_1}~.
    \end{split}
\end{equation}
Now in what follows we only care about extracting the leading $z_{12}^{-1}$ pole. Since $1/P_{12}^2$ $\sim$ $1/{z_{12}}$, the leading term can be obtained by evaluating $\hat{A}_3(z_I)$ and $\hat{A}_{n-1}(z_I)$ at the locus $z_{12}=0$.  In this limit, our parametrizations for the spinors $\lambda_P$ and $\tilde{\lambda}_P$ in~\eqref{eq:p12} simplify to
\be
\omega_P=\omega_1+\omega_2,~~~t_P=\frac{\omega_1 t_1+\omega_2 t_2}{\omega_1+\omega_2},~~~z_P=z_2,~~~\bz_P=\frac{\o_1\bar{z}_1+\o_2\bar{z}_2}{\o_1+\o_2}
\ee
while $\hat{\tilde{\lambda}}_n=\tilde{\lambda}_n$. The $n$-point amplitude then becomes
\begin{equation}
    \begin{split}
        A_n ~=&~ \frac{1}{(-4)z_{12}}\,\sum_{s_P}\,g_{12P}\,2^{s_1+s_2-s_P}\,\bar{z}_{12}^{s_1+s_2-s_P-1}\,\o_1^{s_2-s_P-1}\,\o_2^{s_1-s_P-1}\,
        (\o_1 t_1+\o_2 t_2)^{s_P}\,t_1^{-s_1}\,t_2^{-s_2}\\
        &~\times A_{n-1}^{s_P}\left[\,\o_P,t_P,z_P,\bar{z}_P;\o_3,t_3,z_3,\bar{z}_3;\dots\right] .
    \end{split}
    \label{equ:collinear-splitfunc}
\end{equation}
To evaluate the celestial amplitude we must Mellin transform both sides.  Upon performing the change of variables 
\be
\omega_1=\omega_P \tau,~~~\omega_2=\omega_P(1-\tau),
\ee
implementing a little group rescaling so that $t_1 = t_2 = t$, and series expanding around ${\bz}_P=\bz_2$, we see that the integral factorizes. The $\omega_P$ and $\omega_i$ integrals take the same form as our celestial map~\eqref{eq:mellin} leaving us with
\begin{equation}\label{equ:deriving-OPE}
   \scalemath{.95}{ \badat{3}
        \Tilde{A}_n ~=&~ \frac{1}{-2z_{12}}\,\sum_{J_P}\,g_{12P}\,(2\bar{z}_{12})^{J_1+J_2-J_P-1}\,t^{J_P-J_1-J_2}\,\int_0^{1}d\t\,\t^{\D_1-1+J_2-J_P-1}\,(1-\t)^{\D_2-1+J_1-J_P-1}\\
       &\qquad \sum_{m=0}^{\infty}\,\frac{1}{m!}(\t\bar{z}_{12})^m\,\pa_{\bar{z}_2}^m
        \Tilde{A}_{n-1}^{J_P}\left[\,\D_P,t,z_2,\bar{z}_2;\o_3,t_3,z_3,\bar{z}_3;\dots\right]
\eadat}
\ee
where the conformal dimension $\Delta_P$ is such that $h_P=h_1+h_2- 2.$ The remaining $\tau$ integral evaluates to a Beta function for each term in the sum
  \be\label{ab_bh}
B(a,b)= \int_0^{1}d\t\,\t^{a-1}\,(1-\t)^{b-1},~~~
 a=-\bh_2+\bh_P+m,~~~b=-\bh_1+\bh_P.
 \ee
 Putting this together and recalling our celestial dictionary~\eqref{eq:corr}, we find the leading celestial OPE structure takes the following form
\begin{equation}
    \scalemath{.95}{\badat{3}
        {\cal O}_{\D_1,J_1}(z_1,\bar{z}_1,t)\, {\cal O}_{\D_2,J_2}(z_2,\bar{z}_2,t) ~=&~ \frac{1}{-2z_{12}}\,\sum_{J_P}\sum_{m=0}^{\infty}\,g_{12P}\,\frac{2^{2\bh_P-2\bh_1-2\bh_2}}{m!}\,\bar{z}_{12}^{2\bh_P-2\bh_1-2\bh_2+m}\,t^{2\bh_1+2\bh_2-2\bh_P-1}\\
 & B(-\bh_2+\bh_P+m,-\bh_1+\bh_P)\,\pa_{\bar{z}_2}^m\,{\cal O}_{\D_P,J_P}(z_2,\bar{z}_2,t)
   \eadat}
    \label{equ:leading-OPE}
\end{equation}
where we recall that this expression applies whenever the 3-pt amplitude that contributes is of anti-MHV 
 type, namely $s_1+s_2-s_P>0$ so that~\eqref{3ptsquare} applies.  For the case $s_1+s_2-s_P<0$, the roles of $h_i$ and $\bar{h}_i$ would be flipped as well as the left- and right-handed blocks. Therefore we would consider the $[1,n\rangle$ shift to extract the anti-holomorphic $\zb_{12}^{-1}$ collinear singularity. 
We see that the leading $z_{12}^{-1}$ data includes the full SL($2,\mathbb{R})_{R}$ block~\cite{Guevara:2021abz,Guevara:2021tvr} to all orders in $\bz_{12}$.  
One can similarly include the SL(2,$\mathbb{R})_{L}$ descendants using the global conformal symmetry. Meanwhile the celestial conformal block decomposition~\cite{Nandan:2019jas,Fan:2021isc,Atanasov:2021cje,Fan:2021pbp,Fan:2022vbz,Hu:2022syq} or an application of celestial current algebras~\cite{Banerjee:2020zlg,Banerjee:2020vnt,Banerjee:2021cly,Banerjee:2021dlm} can be used to look for other non-singular contributions to the OPE. 

We close this section by commenting that from the point of view of Feynman's approach, the collinear singularity occurs only when the collinear particles emerge from the same three-point vertex. As such, the splitting function can be also derived solely using Feynman rules \cite{Taylor:2017sph}, which is consistent with, but not predicated upon, the BCFW result.

\subsection{\texorpdfstring{Revisiting the Large-$z$ Behavior}{Revisiting the Large-z Behavior}}
\label{sec:largez}

In the last section, we emphasized that $A^c$ was the only contribution to the collinear singularity and was sufficient for extracting the splitting functions and thus the celestial OPE coefficients.  While we reviewed how to use the BCFW expressions to read off the OPE coefficients, we also emphasized that the splitting function can be completely determined without using the BCFW framework. One can instead try to work in the other direction. As shown in~\cite{Pate:2019lpp}, the leading OPE coefficients are forced by symmetries. We can take this further. If CCFT is like ordinary CFT, the operator spectrum and celestial OPE coefficients should be enough data to construct the full amplitude. As for the celestial block decomposition, the seed for the BCFW recursion relations involves the three-point functions.  However, we also need to know that our shifts are of the allowed type. Here we show how to extract this criterion from the celestial data.

In~\cite{Arkani-Hamed:2008bsc}, Arkani-Hamed and Kaplan reinterpreted the large-$z$ limits of amplitudes in terms of a hard particle propagating with large, complex null momentum, and showed that one could determine the scaling with large-$z$ in a given theory via how this particle coupled to a soft background corresponding to the remaining states.  Indeed,  
very roughly if we look at the large-$z$ behavior under the shift
\be
p_i~\mapsto~ p_i~+~ zq~~~~p_j~\mapsto~ p_j~-~z q
\ee
the two operators are approaching each other on the complexified celestial sphere with energies scaling larger than the remaining scatterers. Combined with the fact that a key part of CCFT involves examining complexified collinear limits to extract the symmetry algebras, we might expect the statement about what shifts are allowed to be encoded neatly in the OPE coefficients. As we will see shortly, this is indeed the case.

Here we show that, in contrast to the naive scaling from Feynman diagrammatics (found by tracking powers of the shifted momenta), the naive scaling expected from the celestial OPE coefficients matches the actual large-$z$ behavior of amplitudes in Yang-Mills theory and Einstein gravity. Our route can be summarized as follows. First we examine the large $z$ limit of the shifted momenta and realize it leads to a coincidence limit on the celestial sphere. The $n$-point shifted correlator then factorizes into two pieces: the shifted OPE coefficients multiplied by an $(n-1)$-point correlator. The large $z$ scaling is captured by the shifted OPE coefficients since the factorized $(n-1)$-point correlator doesn't contribute any $z$-scaling, as explicitly checked. Finally, computing the shifted OPE coefficients (where little group covariance and the fusion rules play important roles) and implementing the inverse Mellin transform yields the large-$z$ scalings for various helicity configurations associated with BCFW shift $\langle J_1,J_2]$ in agreement with the known results~\cite{Arkani-Hamed:2008bsc}. 

The rest of this section is devoted to presenting how this works explicitly. 
Let us first set up the problem. We will consider a $[2,1\rangle$ shift 
\begin{equation}
p'_1 ~=~ p_1 -z\,q ~~,~~ p'_2 ~=~ p_2 + z\,q ~~,~~ q = -|2\rangle[1|~,
\end{equation}
which is equivalent to the following transformation of the celestial variables (assuming for simplicity $\varepsilon_1=\varepsilon_2=1$)
\begin{equation}
    \begin{split}
        \varepsilon_1\,\omega_1 ~\mapsto&~ \varepsilon'_1\,\omega'_1 ~=~ \varepsilon_1\,\omega_1\,\Big( 1-z\,\frac{\varepsilon_2\sqrt{\omega_2}}{\varepsilon_1\sqrt{\omega_1}}\Big) ~=~ \o_1 - z\,\sqrt{\o_1\o_2}  ~~,~~ \\
        z_1~\mapsto&~ z'_1 ~=~ z_1 + \frac{z_{12}\,z}{\frac{\varepsilon_1\sqrt{\omega_1}}{\varepsilon_2\sqrt{\omega_2}} - z} ~~,~~ \bz_1~\mapsto~\bz'_1~=~\bz_1\\
        \varepsilon_2\,\omega_2 ~\mapsto&~ \varepsilon'_2\,\omega'_2 ~=~ \varepsilon_2\,\omega_2\,\Big( 1+z\,\frac{\sqrt{\omega_1}}{\sqrt{\omega_2}}\Big) ~=~ \o_2 + z\,\sqrt{\o_1\o_2} ~~,~~\\
        z_2~\mapsto&~ z'_2 ~=~ z_2  ~~,~~ \bz_2~\mapsto~\bz'_2~=~\bz_2 + \frac{\bz_{12}\,z}{\frac{\sqrt{\omega_2}}{\sqrt{\omega_1}} + z} ~. \\
    \end{split}
    \label{equ:largez-celestialvariable-shift}
\end{equation}
As we have seen in section~\ref{sec:celestial-recursion}, the BCFW shift operation is equivalent to implementing a ``local'' Lorentz transformation acting only on $|\lambda_1\rangle$ and $|\tilde{\lambda}_2]$. Writing 
\begin{equation}
    |\lambda_i'\rangle ~=~ \Lambda_i\,|\lambda_i\rangle ~~,~~ |\tilde{\lambda}'_i] ~=~ \tilde{\Lambda}_i\,|\tilde{\lambda}_i] ~~,~~  \Lambda_i ~=~ \begin{pmatrix}
    d_i & c_i\\
    b_i & a_i\\
    \end{pmatrix} ~~~,~~~
    \tilde{\Lambda} ~=~ \begin{pmatrix}
    \bar{a}_i & -\bar{b}_i\\
    -\bar{c}_i & \bar{d}_i\\
    \end{pmatrix} ~~~,~~~ 
\end{equation}
we want 
\begin{equation}
\begin{split}
   & \Lambda_i ~=~ \mathbb{1}_2~~ (\forall i\ne 1) ~~,~~ \tilde{\Lambda}_j ~=~ \mathbb{1}_2~~ (\forall j\ne 2) ~~,~~  \\
\end{split}
\end{equation}
while
\begin{equation}\label{localLambda}
    \begin{split}
         & |\lambda_1'\rangle~=~\Lambda_1\,|\lambda_1\rangle ~=~ \sqrt{\left|{c_1z_1+d_1}\right|}\,\varepsilon'_1\,\sqrt{2\o'_1 }\begin{pmatrix}
        1\\
        z'_1
        \end{pmatrix} ~~,~~ |\tilde{\lambda}_2'] ~=~\tilde{\Lambda}_2\,|\tilde{\lambda}_2] ~=~ \sqrt{\left|\bar{c}_2\bar{z}_2+\bar{d}_2\right|}\sqrt{2\o'_2}\begin{pmatrix}
        -\bar{z}_2'\\
        1
        \end{pmatrix} ~~,~~\\ 
        & z_1' ~=~ \frac{a_1z_1+b_1}{c_1z_1+d_1} ~~,~~\bz'_2 ~=~ \frac{\bar{a}_2\bz_2+\bar{b}_2}{\bar{c}_2\bz_2+\bar{d}_2}~~,~~
        \o_1' ~=~ |c_1z_1+d_1|\,\o_1 ~~,~~ \o'_2 ~=~ |\bar{c}_2\bz_2+\bar{d}_2|\,\o_2~.
    \end{split}
\end{equation}
Now comparing with (\ref{equ:largez-celestialvariable-shift}), we know that
\begin{equation}
    |c_1z_1+d_1| ~=~ \left|1 - z\frac{\sqrt{\omega_2}}{\sqrt{\omega_1}}\right| ~~,~~ |\bar{c}_2\bz_2+\bar{d}_2|~=~ \left|1 + z\frac{\sqrt{\omega_1}}{\sqrt{\omega_2}}\right|~.
    \label{equ:cz+d}
\end{equation}
Little group covariance plus the Jacobian coming from the $\o$ scaling leads to
\begin{equation}
\begin{split}
     \Lambda_1\,\tilde{\Lambda}_2\,\langle\, {\cal O}_{\D_1,J_1}(z_1,\bz_1)\,&{\cal O}_{\D_2,J_2}(z_2,\bz_2)\,\cdots\rangle \\
     ~=&~ |c_1z_1+d_1|^{-\D_1-J_1}\,|\bar{c}_2\bz_2+\bar{d}_2|^{-\D_2+J_2}\,\langle\,{\cal O}_{\D_1,J_1}(z'_1,\bz'_1)\,{\cal O}_{\D_2,J_2}(z'_2,\bz'_2)\,\cdots\rangle~.
\end{split}
    \label{equ:12transf}
\end{equation}

Now let us look at the $z\to\infty$ limit. Kinematically, the large-$z$ behavior of the $[2,1\rangle$-shift indeed picks out the collinear limit of the shifted momenta, albeit in a complexified direction. Namely, (\ref{equ:largez-celestialvariable-shift}) tells us that $z_{12}$ and $\bz_{12}$ both approach to zero as $z\to\infty$, although $\langle 1'2'\rangle$ remains finite.\footnote{This is because the energies also diverge. Indeed we see that the Mandelstam invariants scale as $s=2p_1\cdot p_2=\langle 1'2\rangle[12']\sim z^0$ and $t=2p_1\cdot p_3=\langle 1'3\rangle[13]\sim z$, which is a crossing image of the standard Regge limit.   As shown in \cite{Fu:2013cza}, the BCFW shift configuration can be described by Regge behavior of the amplitude. From the celestial perspective the Regge limit implies a collinear limit but not vice versa.}
To extract the scaling with large-$z$, the next step is to examine the shifted OPE coefficients. Note that the leading (holomorphic and anti-holomorphic) singularity terms dominate. The leading OPE structures read 
\begin{align}
 {\cal O}_{\Delta_1,J_1}(z_1,\bar{z}_1)\, {\cal O}_{\Delta_2,J_2}(z_2,\bar{z}_2) ~\sim&~ \Bigg[\sum_{J_P}\,\frac{C^{(1)}_{12P}}{z_{12}}\,{\cal O}_{\Delta_P,J_P}(z_2,\bar{z}_2) ~+~ \sum_{J_P}\,\frac{C^{(2)}_{12P}}{\bar{z}_{12}}\,{\cal O}_{\Delta_P,J_P}(z_2,\bar{z}_2)\Bigg]
 \label{equ:leadingOPE}
\end{align}
where
\begin{equation}
    \begin{split}
        C^{(1)}_{12P}~=&~ g_{12P}\, \bar{z}_{12}^{J_1+J_2-J_P-1}\,B(\Delta_1-1+J_2-J_P,~\Delta_2-1+J_1-J_P) \\
        C^{(2)}_{12P} ~=&~ g_{12P}\, z_{12}^{-J_1-J_2+J_P-1}\,B(\Delta_1-1-J_2+J_P,~\Delta_2-1-J_1+J_P)~. \\
    \end{split}
\end{equation}
Acting with the corresponding transformations $\Lambda_1$ and $\tilde{\Lambda}_2$ from~\eqref{localLambda} and applying (\ref{equ:12transf}), we have
\begin{equation}
\begin{split}
    \Lambda_1\,{\cal O}_{\Delta_1,J_1}\tilde{\Lambda}_2\,{\cal O}_{\Delta_2,J_2} ~=~ |c_1z_1+d_1|^{-2h_1}|\bar{c}_2\bar{z}_2+\bar{d}_2|^{-2\bh_2}\,\Bigg[&\sum_{J_P}\,\frac{C^{(1)}_{12P}(z)}{z_{1'2'}}\,{\cal O}_{\Delta_P,J_P}(z'_2,\bar{z}'_2) \\
    &~+~ \sum_{J_P}\,\frac{C^{(2)}_{12P}(z)}{\bar{z}_{1'2'}}\,{\cal O}_{\Delta_P,J_P}(z'_2,\bar{z}'_2)\Bigg]~.
\end{split}
\end{equation}
Then (\ref{equ:largez-celestialvariable-shift}) and (\ref{equ:cz+d}) tell us  that under the $z\to\infty$ limit
\begin{equation}
    c_1z_1+d_1 ~\sim~ z ~~,~~ \bar{c}_2\bar{z}_2+\bar{d}_2~\sim~ z ~~,~~
    z_{1'2'} ~=~ \frac{z_{12}}{1-\frac{\sqrt{\omega_2}}{\sqrt{\omega_1}}\,z} ~\sim~ z^{-1} ~~,~~
\bar{z}'_{12} ~=~ \frac{\bar{z}_{12}}{1+\frac{\sqrt{\omega_1}}{\sqrt{\omega_2}}\,z}  ~\sim~ z^{-1} ~~,~~ 
\end{equation}
while
\begin{equation}
    z'_2 ~\sim~ z_2~~,~~ \bz_2' ~\sim~ \bz_2.
\end{equation}
This implies that the factorized $(n-1)$-point correlator scales as $\mathcal{O}(z^0)$ and indeed we can extract the large-$z$ scaling completely from the OPE coefficients. Finally, we can use an inverse Mellin transform to land on the $z$-scalings of the usual momentum space amplitude. Altogether, we have the following large-$z$ scalings. 
\begin{equation}
    \begin{split}
      &  \text{holomorphic singularity:~} ~\sim~ z^{J_2-J_1-J_P}\\
      &  \text{anti-holomorphic singularity:~} ~\sim~ z^{J_2-J_1+J_P}\\
    \end{split}
    \label{equ:largez-scaling}
\end{equation}
The fusion rules for the little group representations amount to which helicity $J_P$ are involved in the OPEs (i.e. which $g_{12P}$ are non-zero for a given theory).
Some examples comparing the above derivation with known results \cite{Arkani-Hamed:2008bsc} are listed in Table~\ref{tab:largez-bcfw-comparison}. We see that (\ref{equ:largez-scaling}) is in perfect agreement with the expected scaling in each case. 

\begin{table}[htp!]
    \centering
    \begin{tabular}{|c|c|c|c|}
    \hline
      Theory &  $\langle J_1,J_2]$  & $z^{J_2-J_1\pm J_P}$  & expected ${A}_{\rm BCFW}(z)$ \\\hline\hline
    YM  & $\langle +,+]$   & $z^{-1}$ & $z^{-1}$ \\\hline
        & $\langle +,-]$   & $z^{-1}$ & $z^{-1}$ \\\hline
        & $\langle -,+]$   & $z^{3}$  & $z^{3}$ \\\hline
        & $\langle -,-]$   & $z^{-1}$ & $z^{-1}$ \\\hline \hline
    GR  & $\langle ++,++]$   & $z^{-2}$ & $z^{-2}$ \\\hline
        & $\langle --,--]$   & $z^{-2}$ & $z^{-2}$ \\\hline
        & $\langle ++,--]$  & $z^{-2}$  & $z^{-2}$ \\\hline
        & $\langle --,++]$  & $z^{6}$ & $z^{6}$ \\\hline
        & $\langle -+,-+]$   & $z^{2}$ & $z^{2}$ \\\hline
        & $\langle -+,--]$  & $z^{-2}$  & $z^{-2}$ \\\hline
        & $\langle -+,++]$  & $z^{2}$  & $z^{2}$ \\\hline
        & $\langle --,-+]$  & $z^{2}$   & $z^{2}$ \\\hline
        & $\langle ++,-+]$  & $z^{-2}$   & $z^{-2}$ \\\hline\hline
    Photon-Graviton &  $\langle +,+]$   & $z^{0}$ & $z^{0}$ \\\hline
        & $\langle +,-]$  & $z^{0}$  & $z^{0}$ \\\hline
        & $\langle -,+]$   & $z^{4}$  & $z^{4}$ \\\hline
        & $\langle -,-]$  & $z^{0}$ & $z^{0}$ \\\hline 
    \end{tabular}
    \caption{Summary of the large-$z$ behavior for various helicity configurations associated with BCFW shift $\langle J_1, J_2]$ as compared to the results in \cite{Arkani-Hamed:2008bsc}. Note that the 2D spin $J$ is identified with the 4D helicity. 
    }
    \label{tab:largez-bcfw-comparison}
\end{table}

\pagebreak

We close this section with a few remarks. 
\begin{itemize}

\item For YM, the color factors are suppressed  in (\ref{equ:leadingOPE}).  The OPE coefficients will contain the structure constant $f^{abc}$ when we restore the color labels.  The $z$-scalings listed in table~\ref{tab:largez-bcfw-comparison} corresponds to adjacent shifts. Namely, for the color-ordered diagrams where 1 and 2 are adjacent, this collinear configuration matches the above leading collinear singularity.
When 1 and 2 are not adjacent, the $z$-scaling is always smaller than the adjacent case, as discussed in~\cite{Arkani-Hamed:2008bsc}. From (\ref{equ:largez-scaling}) one can also notice that the $z$-scalings in Einstein gravity are the square of the ones in YM, which is one manifestation of the double-copy relations. We will explore this connection in more detail in the upcoming work.

\item As mentioned above, the celestial OPE coefficients are isomorphic to the splitting functions in momentum space. The inverse Mellin transform lets us go in the opposite direction as compared to section~\ref{sec:collinear}. One equivalently derives the large-$z$ scaling in momentum space starting from the splitting functions. 
The scaling of the splitting function is determined by the spin of the exchanged particle ${\rm Split}_{s_1s_2}^{s_3}\sim z^{\pm s_3}$, which underlies the connection to Regge behavior discussed in~\cite{Fu:2013cza}. Meanwhile, the little group factors used to remove the $z$ dependence from the $(n-1)$-point amplitude give rise to an asymmetry between particles 1 and 2 that distinguishes which shifts are allowed.
\item Our main takeaway is as follows. The structure of the leading OPE~(\ref{equ:leadingOPE}) is forced by symmetries \cite{Pate:2019lpp} once a given set of couplings is allowed. The celestial OPE data directly contains the fusion rules by which different helicities are coupled in the three-point interactions. Therefore, the large-z behavior under a given BCFW shift can be inferred from the CCFT. 
\end{itemize}

\section{MHV Gluon Sector: BG Equations and Beyond}\label{sec:BG}

While we saw that the collinear and large-$z$ limits could be extracted for any amplitude, our derivation of the tower of soft currents in section~\ref{sec:soft_currents} assumed that we could restrict ourselves to the collinear factorization channel.  That tower was interpreted as a $w_{1+\infty}$ symmetry in~\cite{Strominger:2021mtt}. Such an abundance of symmetries is tantalizing for a bootstrap program.  In this section we will focus on the MHV sector where $A^c$ contains all the data of the amplitude and we can more readily exploit these symmetries.

 Recent progress has been made in this endeavor. Celestial amplitudes in the MHV sector are governed by a set of PDEs identified by Banerjee, Ghosh, and collaborators~\cite{Banerjee:2020kaa,Banerjee:2020vnt}.  An attempt to classify solutions of these PDEs yields an alternate set of amplitudes to the standard Mellin-transformed ones, pointing to interesting generalizations of this framework~\cite{Fan:2022vbz}.  In this section, we explore how these equations follow from the BCFW recursion relations~\cite{Hu:2021lrx} and generalize this to an infinite tower of PDEs generated by the BG equations. We realize that the BCFW shift operation dressed with the gauge transformations can be recast as global transformations plus soft gluon insertions. This interpretation can be straightforwardly generalized to the super-BCFW shift acting on the superamplitude and, consequently, the supercharge and soft gluino mode insertions appear in a supersymmetric version of the null state.

\subsection{A Tower of PDEs}\label{sec:mhvpde}

In \cite{Banerjee:2020vnt}, Banerjee and Ghosh derived the first set of gluon BG equations by comparing soft and collinear limits of celestial amplitudes.  Namely, they looked at the SL($2,\mathbb{R})_L$ block structure and considered two limits, in turn:  (a) starting from the soft theorem, taking $\bz_{12}=0$ and expanding $z_{12}$; (b) starting from the OPE expansion, taking $\bz_{12}=0$, and then calculating the soft current. So long as soft and collinear limits commute, (a) and (b) should give the same result.

The authors of~\cite{Banerjee:2020vnt} focus on the SL($2,\mathbb{R})_L$ block for $\Delta=0$. At level-0, the subleading soft theorem and OPE expansion commute automatically. Meanwhile, we get a first-order PDE when we try to match terms at level-1. The first aim of this section is to consider what happens when we continue this procedure to higher orders, looking at the SL$(2,\mathbb{R})_L$ descendants for the full tower of soft theorems. We will do so by taking a step back to see how the BG equations arise from the BCFW recursion relation.  Namely, we know how a finite BCFW shift works in the collinear channel, we know how to recast the soft/collinear commutator in terms of this shift, and we know how to extend this to all orders in the shift parameter $z$. Because the soft current algebra can be generated by the first few modes (up to subleading order in gauge theory, sub-subleading order in gravity), we expect the same should be true for this tower of PDEs.  We verify that this is the case.  In \cite{Banerjee:2020vnt}, the resulting differential equation could, conversely, be used to determine the leading term in the celestial gluon OPE, given a suitable ansatz.  This is consistent with the recursion procedure in~\cite{Pate:2019lpp}.  We show how this is reproduced in our setup in appendix~\ref{appen:softtocollinear}. 

In section~\ref{sec:SClim} we showed that both the soft and collinear limits followed from the BCFW recursion relation. Only the $A^c$ term was necessary for the collinear limits.  Because this is the only contribution to the MHV amplitudes, we expect to have a lot more control in this sector. Here we will show that exponentiated form of the BCFW shift relations implies that the BG equations exponentiate to a tower of constraints.\footnote{In appendix~\ref{appen:BGfromParkeTaylor}, we show how this PDE tower follows straightforwardly if we start from the Parke-Taylor amplitude. Since the BCFW recursion relations can be used to derive the Parke-Taylor amplitudes, these two routes are equivalent.  Because our emphasis here is on interpreting BCFW in the celestial context, we want to focus on a construction that is more readily generalized beyond the MHV sector.}

Let us start with the finite $\langle i,i-1]$ BCFW shift\footnote{A discussion of $\langle i,j]$ BCFW shifts where $i$ and $j$ are not adjacent can be found in appendix \ref{appen:Dij-PDEs}.}
 \begin{equation}
     \exp\,\Big(z\, D_{i,i-1}\Big) \, {A} ~=~ \hat{A}(z)
     \label{equ:exp-z}
 \end{equation}
 where $D_{i,i-1}$ defined in (\ref{eq:dij}), and consider what happens when we expand around small $z$.  If we know the full form of the right-hand side, we end up with a tower of differential constraints. This is the case for tree-level color-ordered MHV amplitudes. 
 For this particular shift $\hat{A}(z)$ only has one simple pole at $z_I=-\frac{\langle i,i+1 \rangle}{\langle i-1,i+1\rangle}$.\footnote{This shift is valid for $(s_i,s_{i-1})=(+,\pm),(-,-)$ as we've discussed in section~\ref{sec:largez}.
 }
Since $\hat{A}(z)$ is regular at $z=0$, its Laurent series reduces to a Taylor series of the form 
\begin{equation}
    \hat{A}(z) ~=~ \sum_{p=0}^{\infty}\,a_p\,z^p ~=~ \sum_{p=0}^{\infty}\,\frac{a_p}{\hat{A}(0)}\,z^p\,\hat{A}(0)
    \label{equ:Taylor-series-A(z)}
\end{equation}
where
\begin{equation}
\begin{split}
    a_p~=&~ \frac{1}{2\pi i}\oint_{\gamma}\,\frac{\hat{A}(z)}{z^{p+1}}\,dz 
    ~=~ -{\rm Res}_{z\to z_I}\,\frac{\hat{A}(z)}{z^{p+1}}
    ~=~ -\,\frac{{\rm Res}_{z\to z_I}\, \hat{A}(z)}{z_I^{p+1}}
\end{split}
\end{equation}
and $\gamma$ is a small contour around the point $z=0$. Writing the un-shifted amplitude $A$ as
\begin{equation}
   A ~=~  \hat{A}(0)~=~ \frac{1}{2\pi i}\oint_{\gamma}\,\frac{\hat{A}(z)}{z}\,dz  ~=~ -\,{\rm Res}_{z\to z_I}\,\frac{\hat{A}(z)}{z} ~=~ -\,\frac{{\rm Res}_{z\to z_I}\,\hat{A}(z)}{z_I}
\end{equation}
we find
\begin{equation}
    a_p ~=~ \frac{\hat{A}(0)}{z_I^p} ~~~\rightarrow~~~ \hat{A}(z) ~=~ A\,\sum_{p=0}^{\infty}\,z^p\,(-1)^p\,\left(\frac{\langle i-1,i+1\rangle}{\langle i,i+1 \rangle}\right)^p.
\end{equation}
Series expanding (\ref{equ:exp-z}) we thus have
\begin{equation}
    \sum_{p=0}^{\infty}\,\frac{1}{p!}\,z^p\,(D_{i,i-1})^p\, A_n[123\cdots n] ~=~ \sum_{p=0}^{\infty}\,(-1)^p\,z^p\,\left(\frac{\langle i+1,i-1 \rangle}{\langle i+1,i \rangle}\right)^p
    \,A_n[123\cdots n].
   \label{equ:PT-PDE}
\end{equation}
As explored further in appendix~\ref{appen:BGfromParkeTaylor}, the Parke-Taylor denominator implies this simple form for the action of $D_{i,i-1}$ on color-ordered MHV amplitudes.

Let us now see what happens when we map this relation to the celestial sphere.  Using \eqref{equ:Dij-CS}, and setting $t_i=t_j$ so that the celestial OPE factorizes, we find that at $\mathcal{O}(z^p)$ we have 
\begin{equation}
    \begin{split}
        \Bigg\{\,\Big[\,-\, \Omega_{i,i-1}\,&\left(\,2h_i+z_{i,i-1}\,\pa_i \,\right) + \Tilde{\Omega}_{i,i-1}\,\left(\,2\Bar{h}_{i-1}+\Bar{z}_{i-1,i}\,\Bar{\pa}_{i-1} \,\right)\,\Big]^p \\
        &~-~ (-1)^{p}\,p!\,\Omega_{i,i-1}^p\,\sum_{l=0}^p\,\left(\frac{z_{i+1,i-1}}{z_{i+1,i}}\right)^p
        \,\Bigg\}\,\langle {\cal O}^{\epsilon_1}_{\D_1,J_1}\cdots{\cal O}^{\epsilon_i}_{\D_i,+}\cdots{\cal O}^{\epsilon_n}_{\D_n,J_n} \rangle ~=~ 0
    \end{split}
\end{equation}
where $\Omega_{i,i-1}$ and $\Tilde{\Omega}_{i,i-1}$ are given by \eqref{LambdatLambda} for $j=i-1$.   The case $p=0$ is trivial.  Meanwhile, at $p=1$, we reproduce the color-ordered Banerjee-Ghosh equation
\begin{equation}
    \begin{split}
       \left(\,\pa_i + \frac{\Delta_i}{z_{i,i-1}} - \frac{1}{z_{i+1,i}} ~+~ \epsilon_i\epsilon_{i-1}\,\frac{2\bar{h}_{i-1}-1+\Bar{z}_{i-1,i}\Bar{\pa}_{i-1}}{z_{i-1,i}}\,T_i\,T^{-1}_{i-1}\,\right) \Tilde{A}~=~0.
    \end{split}
    \label{equ:color-ordered-BG}
\end{equation}
This has a natural interpretation as a relation amongst Poincar\'e descendants since the weight shifting operators $T_i$ appear in the action of the translation generators. Note that as a consistency check we can derive this equation for the $i=1$ case starting from~\eqref{equ:c-recursion-2}.  Because the dependence on $z_1$ is solely in the exponentiated prefactor, one can take appropriate derivatives of both sides and match the BG equation form. 

From the structure of~\eqref{equ:exp-z} we see that this constraint exponentiates so that the tower of PDEs implied by the BCFW recursion relation follows from the $p=1$ term. While the fact that we only need the $p=1$ term to determine the $p>1$ tower means we don't get additional constraints, the nature of this exponentiation implies that the higher-order terms in the celestial OPE corresponding to tree-level MHV amplitudes also have this structure.

\subsection{BCFW as Soft Insertions}\label{sec:modifiedBCFWtoBG}
In section~\ref{sec:mhvpde}, the discussion focused on the color-ordered story, which suggested that the BG equation basically comes from the BCFW shift.  Because the BG equation can be phrased in terms of null states involving soft currents, one might expect the BCFW shift can be related to soft gluon insertions.  
To have a concrete statement about such a relation between BCFW shift and soft mode insertions, we would like to extend the discussion to study the full, color-dressed, amplitudes/celestial correlators. 
In practice, we can promote the on-shell data $\{|i\rangle,|i]\}$ to $\{|i,a_i\rangle,|i,a_i]\}$ to restore the gauge group factors. While we've considered the action of Lorentz transformations on each external state above, this will make it easier for us to take into account any gauge transformations when 
In this section, we show that the finite BCFW shift operations, accompanied with particular gauge transformations, can be recast in terms of certain Poincar\'e descendants and soft gluon insertions. A quick application is to consider the MHV tree-level example, where the Parke-Taylor denominator hands us the BG null state expression. 

Consider a generalized $\langle i,k]$ BCFW shift accompanied by a gauge transformation of the form
\begin{equation}
\begin{split}
   & \begin{cases}
    |i,a_i\rangle ~\mapsto~ |i,a_i\rangle + z\,\sum_{k\ne i}\,\frac{(-2)\omega_i}{\langle ik \rangle}\,{\cal T}_k^a\,|k,a_k\rangle \\
    |k,a_k] ~\mapsto~ |k,a_k] - z\,\frac{(-2)\omega_i}{\langle ik \rangle}\,{\cal T}_k^a\,|i,a_i]~~~\text{for all }k\ne i~.\\
    \end{cases}\\
\end{split}
\label{equ:generalized-shift}
\end{equation}
The rescaling factor $\frac{(-2)\omega_i}{\langle ik \rangle}$ will appear in appendix \ref{appen:BCFW-ASG} when we try to relate the action of a BCFW shift to that of the asymptotic symmetry group. The operation (\ref{equ:generalized-shift}) can be implemented on an amplitude (and the corresponding celestial correlator) by the following exponential
\begin{equation}\resizebox{0.9\textwidth}{!}{$%
     \exp{\left\{z\,\Big[\,-j_0^a(i)\,L_{-1}(i) + 2\,L_0(i)\,j_{-1}^a(i)
          + P_{-\frac{1}{2},-\frac{1}{2}}(i)\,{\cal J}^a_{-\frac{1}{2},\frac{1}{2}}(i) \Big]\right\}}\,\Big\langle {\cal O}^{a_i}_{\Delta_i,J_i}(z_i,\bar{z}_i)\prod_j{\cal O}^{a_j}_{\Delta_j,J_j}(z_j,\bar{z}_j) \Big\rangle$}%
          \label{equ:generalized-shift-operator}
\end{equation}
where the actions of these global transformations and soft gluon insertions are listed below
\begin{align}
    j_0^a(i) ~=&~ -\,{\cal T}_i^a ~~,~~  L_0(i) ~=~ h_i ~~,~~ P_{-\frac{1}{2},-\frac{1}{2}}(i){\cal O}^{a_i}_{\Delta_i,J_i} ~=~ {\cal O}^{a_i}_{\Delta_i+1,J_i} \\
    j_{-1}^a(i) ~=&~  \sum_{k\ne i}\,\frac{{\cal T}_k^a}{z_{ki}}~~,~~
    {\cal J}_{-\frac{1}{2},\frac{1}{2}}^a(i) ~=~ \sum_{k\ne i}\,\varepsilon_i\varepsilon_k\,\frac{2\bar{h}_k-1+\bar{z}_{ki}\bar{\partial}_k}{z_{ik}}\,T_k^{-1}\,{\cal T}_k^a~.
\end{align}
Note that the soft current insertions $j(i)$ and ${\cal J}(i)$ means that we are adding a soft gluon mode that is collinear with ${\cal O}_i$. The Ward identity (\ref{wardid}) tells us that inserting a soft operator is equivalent to doing transformations on all hard operators. Here we are saying the same thing from the opposite direction. Namely, from (\ref{equ:generalized-shift}) to (\ref{equ:generalized-shift-operator}), we are implementing transformations on the hard operators and recasting them as soft insertions using the soft theorems. 

The exponential nature of (\ref{equ:generalized-shift-operator}) leads us to focus on the first-order story. Restoring gauge group factors to our discussion of the MHV gluon example from section~\ref{sec:mhvpde}, we can play a similar game and find
\begin{equation}
\begin{split}\label{BGnull}
    \Bigg[\,-\,j_0^a(i)\,L_{-1}(i) + (2\,L_0(i)-1)\,j_{-1}^a(i)
          &+ P_{-1,-1}(i)\,{\cal J}^a_{-\frac{1}{2},\frac{1}{2}}(i)  \Bigg]\,\Big\langle {\cal O}^{a_i}_{\Delta_i,J_i}\prod_j{\cal O}^{a_j}_{\Delta_j,J_j} \Big\rangle_{MHV}\\
       & - j_{-1}^{a_i}(i)\,\Big\langle {\cal O}^{a}_{\Delta_i,J_i}\prod_j{\cal O}^{a_j}_{\Delta_j,J_j} \Big\rangle_{MHV}  = 0
          \end{split}
\end{equation}
which yields the BG null state for the MHV gluon sector presented in \cite{Banerjee:2020vnt}.

\subsection{From Super-BCFW to a Modified Null State}\label{sec:superBCFW}
The soft insertion interpretation of the BCFW shift can be straightforwardly generalized to the super-BCFW shift. In~\cite{Hu:2021lrx} the color-ordered BG equation was generalized to MHV superamplitudes. Here we will extend this result by restoring the color factors and identifying a null state for the superamplitude, following a procedure analogous to the steps that took us from~\eqref{equ:color-ordered-BG} to~\eqref{BGnull}.
When we extend the global symmetry group from Poincar\'e to super-Poincar\'e, we need to implement a super-BCFW shift that respects both momentum conservation and supercharge conservation. Shifting only $\lambda$ and $\tilde{\lambda}$ is not enough; the Grassmann-valued coordinates have to be shifted appropriately as well. We can write a super-BCFW $\langle i,j]$-shift as follows
\begin{equation}
    \lambda_i ~\mapsto~ \lambda_i + z\,\lambda_j ~~,~~ \tilde{\lambda}_j ~\mapsto~ \tilde{\lambda}_j - z\,\tilde{\lambda}_i ~~,~~\eta_j^A ~\mapsto~ \eta_j^A - z\,\eta_i^A ~.
\end{equation}

In analogy to (\ref{equ:generalized-shift}), one can write a generalized super-BCFW shift and explore the effect of this shift operation on the superamplitude. Including the Grassmann coordinate shifts, we have
\begin{equation}
\begin{split}
   & \begin{cases}
    |i,a_i\rangle ~\mapsto~ |i,a_i\rangle + z\,\sum_{k\ne i}\,\frac{(-2)\omega_i}{\langle ik \rangle}\,{\cal T}_k^a\,|k,a_k\rangle \\
    |k,a_k] ~\mapsto~ |k,a_k] - z\,\frac{(-2)\omega_i}{\langle ik \rangle}\,{\cal T}_k^a\,|i,a_i]~~~\text{for all }k\ne i\\
    (\eta_k^A,a_k) ~\mapsto~ (\eta_k^A,a_k) - z\,\frac{(-2)\omega_i}{\langle ik \rangle}\,{\cal T}_k^a\,(\eta_i^A,a_i)~~~\text{for all }k\ne i\\
    \end{cases}\\
\end{split}
\label{equ:generalized-super-shift}
\end{equation}
The operation (\ref{equ:generalized-super-shift}) can be implemented on a superamplitude as follows
\begin{equation}\resizebox{0.9\textwidth}{!}{$
    \begin{aligned}
        \exp\,z\,\left\{\, -j_0^a(i)L_{-1}(i) +2h_i\,j^a_{-1}(i) +P_{-\frac{1}{2},-\frac{1}{2}}{\cal J}_{-\frac{1}{2},\frac{1}{2}}^a(i) - \frac{1}{\sqrt{2}}\Tilde{Q}_{-\frac{1}{4},-\frac{1}{4}}^{A}(i)\, S^a_{-\frac{3}{4},\frac{1}{4}}(i)    \right \}\,\Big\langle \prod_j \Omega^{a_j}_j \Big\rangle
    \end{aligned}$}
    \label{equ:BG-superamplitude}
\end{equation}
where 
\begin{equation}
    \Tilde{Q}_{-\frac{1}{4},-\frac{1}{4}}^{A}(i) ~=~ \varepsilon_i\,\sqrt{2}\,T_i^{\frac{1}{2}}\,\eta_i^A
\end{equation}
is the $i^{\rm th}$ supercharge and 
\begin{equation}
     S^a_{-\frac{3}{4},\frac{1}{4}}(i) ~=~ \sum_{k\ne i}\varepsilon_k T_k^{-\frac{1}{2}}\frac{(-1)^{\sigma_k}{\cal T}_k^a}{z_{ik}}\,\frac{\partial}{\partial \eta_k^A}
     \label{equ:leading-soft-gluino}
\end{equation}
amounts to inserting a leading soft ($\D\to \frac{1}{2}$) helicity $+\frac{1}{2}$ gluino that is collinear to ${\cal O}_i$~\cite{Fotopoulos:2020bqj}. Note that if we perform an $\eta$-expansion of the super-correlator to look at the component correlators, we see that the $\sigma_k$ appearing in~\eqref{equ:leading-soft-gluino}  is the number of fermions inserted to the left of ${\cal O}_k$ and the $\frac{\partial}{\partial \eta_k^A}$ operator lowers the helicity of ${\cal O}_k$ via $J_k\mapsto J_k-\frac{1}{2}$.
As before, consider the first-order equation and the form of the Parke-Taylor denominator imply
\begin{equation}
    \begin{split}
       &\Bigg[ -j_0^a(i)L_{-1}(i)\,\delta^{a_i,b} + (2h_i-1)\,j^a_{-1}(i)\,\delta^{a_i,b} +P_{-\frac{1}{2},-\frac{1}{2}}{\cal J}_{-\frac{1}{2},\frac{1}{2}}^a(i)\,\delta^{a_i,b} - j_{-1}^{a_i}(i)\,\delta^{a,b}\\ 
      & ~-~ \frac{1}{\sqrt{2}}\Tilde{Q}_{-\frac{1}{4},-\frac{1}{4}}^{A}(i)\, S^a_{-\frac{3}{4},\frac{1}{4}}(i)\,\delta^{a_i,b}\,\Bigg]\,\Big\langle\Omega^{b}_i \prod_j \Omega^{a_j}_j \Big\rangle ~=~ 0
    \end{split}
    \label{equ:SYM-BG}
\end{equation}
which yields the null states for the MHV sector in ${\cal N}=4$ SYM. 

Note that in (\ref{equ:SYM-BG}) $i$ can be arbitrary. Previously in the pure gluon case, we restrict $J_i=+1$ and only have null states associated with positive helicity gluon operators. One can see that (\ref{equ:SYM-BG}) reduces to the same result by looking at the component correlator accompanying with $(\eta_s)^4(\eta_t)^4$ where $i\ne s,t$. The existence of $\Tilde{Q}^A(i)$ means that the original BG operation actually raises the helicity of ${\cal O}_i$. To give another precise example, let's consider $J_i=-1$, pick up the components accompanying with $(\eta_i)^4(\eta_s)^4$, and we have 
\begin{equation}\resizebox{0.9\textwidth}{!}{$%
    \begin{aligned}
       &\Bigg[ -j_0^a(i)L_{-1}(i)\,\delta^{a_i,b} + (2h_i-1)\,j^a_{-1}(i)\,\delta^{a_i,b} +P_{-\frac{1}{2},-\frac{1}{2}}{\cal J}_{-\frac{1}{2},\frac{1}{2}}^a(i)\,\delta^{a_i,b} - j_{-1}^{a_i}(i)\,\delta^{a,b} \Bigg]\Big\langle{\cal O}^{b}_{i,-}{\cal O}^{a_s}_{s,-} \prod_j {\cal O}^{a_j}_{j,+} \Big\rangle \\
      &  ~=~ \sum_{k\ne i,s}\,\varepsilon_i\varepsilon_k\,T_i^{\frac{1}{2}}T_k^{-\frac{1}{2}}\,\frac{{\cal T}_k^a}{z_{ik}}~\Big\langle \bar{\lambda}^{a_i}_{i,A} \lambda^{a_k,A}_{k} \prod_j {\cal O}^{a_j}_j \Big\rangle
    \end{aligned}$}
    \label{equ:negative-helicity-gluon-BG}
\end{equation}
Namely, the original BG operation (LHS in the equation above) raises the helicity $-1$ gluon ${\cal O}_{i,-}$ to helicity $-\frac{1}{2}$ gluino $\bar{\lambda}_{i,A}$.

\section{Discussion}\label{sec:disccussion}

We will close with a recap of the highlights from each section, emphasizing what was reviewed and what was new, how these components fit together, and where interesting avenues for future work appear.

\paragraph{Shifted Kinematics and Asymptotic Symmetries}
Section~\ref{sec:BCFW} started with a review of the BCFW recursion relation and its celestial incarnation, following~\cite{Elvang:2013cua} and~\cite{Guevara:2019ypd}, respectively. The new take here was to lean into the interpretation of the spinor shifts in terms of asymptotic symmetry generators, which we later applied to the BG null state analysis in section~\ref{sec:modifiedBCFWtoBG}.  Details of the relation between the infinitesimal complex shifts of the external kinematics and superrotations are fleshed out in appendix~\ref{app:zshiftASG}.

A natural extension of this discussion is to ask whether or not it is natural to exponentiate this statement.  Namely, to what extend do we want to equate implementing finite shifts of the hard particles with insertions of a particular cloud of soft gravitons? The coupling to the soft graviton implies the quasi-primaries are promoted to Virasoro primaries
\be\label{virop}
{\cal O}^{\epsilon'}_{h,\bar h}(z',\bz')=\left|\frac{\p z}{\p z'}\right|^h\left|\frac{\p \bz}{\p \bz'}\right|^{\bar h}{\cal O}^\epsilon_{h,\bar h}(z,\bz)~.
\ee
Besides this action on the hard operators, finite superrotations induce a Schwarzian shift in the metric, changing the superrotation vacuum~\cite{Strominger:2016wns,Compere:2016jwb,Adjei:2019tuj,Pasterski:2022lsl}.

\paragraph{CCFT Data and BCFW}
In section~\ref{sec:SClim} we started by reviewing how to extract soft and collinear limits of scattering, following~\cite{Guevara:2019ypd} and~\cite{Pate:2019lpp}. The point of celestial holography is to apply CFT bootstrap techniques to the celestial program, and it is worth highlighting three interrelated features that distinguish this setup from how one would treat ordinary 2D CFTs.
\begin{enumerate} 
\item Each bulk field corresponds to a continuum of 2D operators with tuneable conformal dimension~\cite{deBoer:2003vf,Pasterski:2017kqt}.
\item The currents that we get by continuing to special values of the conformal dimension give constraints on the OPE coefficients that can be sufficient to bootstrap them~\cite{Pate:2019lpp}.
\item The OPE data encodes enhancements to the symmetry algebra~\cite{Guevara:2021abz}.
\end{enumerate}
While we discussed how to arrive at the CCFT data from the BCFW recursion results, a recurring theme of our investigations has been to try to work both ways. This is highlighted in section~\ref{sec:largez}, we showed how to read off the large-$z$ behavior, and thus determine when a given BCFW shift is allowed, starting from the celestial OPE data.

\paragraph{Factorization Channels vs Conformal Projectors}
The seed data is the three-point function for both the celestial and BCFW recursion relations, however we want to emphasize that the way that these recursive procedures organize amplitudes is quite different.  BCFW takes us from lower point amplitudes to higher point amplitudes at complexified kinematics where intermediate states go on-shell. These amplitudes should match the connected part of celestial correlators, which we expect to be able to recast in terms of a 2D conformal block decomposition~\cite{Nandan:2019jas,Fan:2021isc,Atanasov:2021cje,Fan:2021pbp,Fan:2022vbz,Hu:2022syq}. 

The first examination of the celestial incarnation of BCFW~\cite{Pasterski:2017ylz} was indeed aimed at exploring such a connection, using the tree level 4-gluon amplitude as an example.  We've summarized the highlights of that computation in appendix~\ref{appen:4gluon} for convenience. The key technical difficulty is in restoring the momentum conservation delta functions rather than working with the stripped amplitudes. However, one can massage the factorized terms in~\eqref{BCFWAn} back into the form 
\begin{equation}
\begin{split}
   {\cal A}_n ~
      =&~ -\,\sum_{I}\, {\rm Res}_{z_I} z^{-1}\int \frac{d^4 P }{P^2}\hat{\cal A}_{L}[\hat{i}^{s_i},\,-P](z)\hat{\cal A}_{R}[P,\,\hat{j}^{s_j}](z) \\
\end{split}
\label{equ:alr}
\end{equation}
 via steps that amount to essentially backtracking the standard steps from the Feynman diagram whose internal line is going on-shell to the factorized product of on-shell amplitudes, so that we have an object whose celestial transform we know. While the structure of the BCFW and CFT recursions differ at higher point, in the 4-point example the expression for the BCFW recursion only involves a single factorization channel, deceptively matching the structure of the 2D CFT iteration where we'd expect
\cite{SimmonsDuffin:2012uy}
\begin{equation}
    g_{\O_P} + {\rm ``shadow~block"}~\sim~ \int\,dz_P\,d\bz_P\,\langle\,\O_1\O_2\O_P\rangle\,\langle\,\Tilde{\O}_P\O_3\O_4\rangle
\end{equation}
where $\Tilde{\O}_P$ is the shadow transform of $\O_P$. Namely, assuming an isomorphism between the 4D and 2D Hilbert spaces, and a state operator correspondence, it is tempting to match the BCFW expression to an exchange mediated by operators corresponding to single-particle external states.

\paragraph{Beyond the MHV Sector}

In section~\ref{sec:BG} we focused on the MHV sector, where we could take advantage of our expressions from section~\ref{sec:soft_currents} that relied on our restriction to contributions from $A^c$. In section \ref{sec:mhvpde} we saw how the BG equations arose from demanding soft and collinear limits commute. The question of whether this continues to hold is important for understanding the full celestial symmetry algebra. Starting from (\ref{equ:exp-z}) and the analytic structure of the shifted amplitude $\hat{A}(z)$, one can straightforwardly reproduce the color-ordered BG equation.  This procedure suggests a general way to construct the differential constraints for tree-level amplitudes, whose celestial avatars are candidate null states.  Note that the story here only depended on the analytic structure of $\hat{A}(z)$, namely the positions of all poles, the orders of the poles, and the fall-off condition at infinity, so one can extend the discussion to NMHV and higher orders. One would expect that for non-MHV cases, there are more poles, more non-vanishing channels, and consequently higher-degree of polynomials in the numerator of $\hat{A}(z)$, which result in an analog of the BG equation taking a more complicated form.

\section*{Acknowledgements}
We would like to thank Shamik Banerjee, Justin Kulp, Luke Lippstreu, Lecheng Ren, Atul Sharma, Marcus Spradlin, Akshay Yelleshpur Srikant, and Anastasia Volovich for useful conversations. The research of YH has been supported in part by the endowment from the Ford Foundation Professorship of Physics and the Physics Dissertation Fellowship provided by the Department of Physics at Brown University. YH also acknowledges the support of the Brown Theoretical Physics Center.  The research of SP is supported by the Celestial Holography Initiative at the Perimeter Institute for Theoretical Physics and has been supported by the Sam B. Treiman Fellowship at the Princeton Center for Theoretical Science. Research at the Perimeter Institute is supported by the Government of Canada through the Department of Innovation, Science and Industry Canada and by the Province of Ontario through the Ministry of Colleges and Universities.

\appendix
\section{Conventions}\label{app:conventions}
 In this appendix we will briefly summarize our conventions, which follow from those of~\cite{Elvang:2013cua}.  Throughout this paper, we go between four-vector and bi-spinor notation via
\be
v_{\alpha{\dot \beta}}=v_\mu(\sigma^\mu)_{\alpha\dot \beta}\,,~~~v^\mu=-\frac{1}{2}\mathrm{tr}(v\bar{\sigma}^\mu)\,,
\ee
where
\be\label{ivdw}
(\sigma^\mu)_{\alpha\dot \beta}=(\mathds{1},\sigma^i)_{\alpha \dot \beta}\,,~~(\bar{\sigma}^\mu)^{{\dot \alpha} \beta}=(\mathds{1},-\sigma^i)^{{\dot \alpha} \beta}\,,
\ee
and undotted (dotted) indices are raised and lowered with
\be\label{lc2}
\varepsilon^{\alpha \beta}=\varepsilon^{\dot{\alpha}\dot{\beta}}=-\varepsilon_{\alpha\beta}=-\varepsilon_{\dot{\alpha}\dot{\beta}}=\left(\begin{array}{cc}
0&1\\-1&0\\
\end{array}\right).
\ee
 In spinor-helicity notation the momentum~\eqref{eq:pi} becomes 
\be
p_i^{\alpha\dot\alpha}=-\,\lambda^{\alpha}_i\tilde{\lambda}^{\dot\alpha}_i
\ee
where
\begin{equation}
  \badat{3}
        \lambda_i^{\a} ~=&~ |\lambda_i\rangle^{\a} ~=~ \epsilon_i\,\sqrt{2\omega_it_i}\,\begin{pmatrix}
        1\\
        z_i
        \end{pmatrix} ~~,~~
        &\lambda_i{}_{\a} ~=&~ \langle\lambda_i|_{\a} ~=~ \epsilon_i\,\sqrt{2\omega_it_i}\,\begin{pmatrix}
        -z_i \\
        1
        \end{pmatrix} ~~,\\
        \Tilde{\lambda}_{i,\Dot\a} ~=&~ |\Tilde{\lambda}_i]_{\Dot\a} ~=~ \sqrt{2\omega_it^{-1}_i}\,\begin{pmatrix}
        -\Bar{z}_i\\
        1
        \end{pmatrix} ~~,~~
        &\Tilde{\lambda}_{i}^{\Dot\a} ~=&~ [\Tilde{\lambda}_i|^{\Dot\a} ~=~ \sqrt{2\omega_it^{-1}_i}\,\begin{pmatrix}
        1\\
        \Bar{z}_i
        \end{pmatrix} ~~,~~ \\
         \langle ij \rangle ~=&~ (-2)\,\epsilon_i\epsilon_j\,\sqrt{\o_i\o_jt_it_j}\,z_{ij} ~~,~~ & [ij] ~=&~ 2\,\sqrt{\o_i\o_jt_i^{-1}t_j^{-1}}\,\bz_{ij}~.
\eadat
    \label{equ:spinorhelicity-celestial}
\end{equation}
Here we have restored a little group rescaling factor $t_i$, which will help us evaluate derivatives with respect to the spinor coordinates. Using the chain rule we can derive the map between the derivatives respect to $\{\lambda,\Tilde{\lambda}\}$ and $\{z_i,\zb_i,t_i,\omega_i\}$
\begin{equation}\scalemath{0.96}{
    \badat{3}
        \frac{\pa}{\pa z_i} ~=&
        ~ \lambda_i^1\,\frac{\pa}{\pa \lambda_i^2}~~,~~ 
        &~~~ \frac{\pa}{\pa \lambda_i^1} ~=&~ \frac{1}{\epsilon_i \sqrt{2\omega_i t_i}}  \left( \omega_i \frac{\pa}{\pa \omega_i} ~+~ t_i\frac{\pa}{\pa t_i} ~-~ z_i \frac{\pa}{\pa z_i} \right)~~,~~\\
        \frac{\pa}{\pa \Bar{z}_i} ~=&
        ~ -\,\Tilde{\lambda}_{i,2}\,\frac{\pa}{\pa \Tilde{\lambda}_{i,1}}~~,~~ 
        &~~~ 
            \frac{\pa}{\pa \lambda_i^2} ~=&~ \frac{1}{\epsilon_i \sqrt{2\omega_i t_i}}\frac{\pa}{\pa z_i}~~,~~ \\
     \omega_i \frac{\pa}{\pa \omega_i} ~=&~ \frac{1}{2}\,\lambda_i^{\Dot\a}\frac{\pa}{\pa \lambda_i^{\Dot\a}} ~+~ \frac{1}{2}\,\Tilde{\lambda}_{i,\a}\frac{\pa}{\pa \Tilde{\lambda}_{i,\a}}~~,~~ 
     &~~~    \frac{\pa}{\pa \tilde{\lambda}_{i,\dot{1}}} ~=&~ -\,\sqrt{\frac{t_i}{2\omega_i}}\frac{\pa }{\pa \bar{z}_i}~~,~~ \\
      t_i\frac{\pa}{\pa t_i} ~=&~ \frac{1}{2}\,\lambda_i^{\Dot\a}\frac{\pa}{\pa \lambda_i^{\Dot\a}} ~-~ \frac{1}{2}\,\Tilde{\lambda}_{i,\a}\frac{\pa}{\pa \Tilde{\lambda}_{i,\a}}~~,~~ 
      &~~~  \frac{\pa}{\pa \tilde{\lambda}_{i,\dot{2}}} ~=&~ \sqrt{\frac{t_i}{2\omega_i}}   \left( \omega_i \frac{\pa}{\pa \omega_i} ~-~ t_i\frac{\pa}{\pa t_i} ~-~ \bar{z}_i \frac{\pa}{\pa \bar{z}_i} \right)~.~~
      \eadat}
      \label{equ:derivtoCS}
\end{equation}
When we go to the celestial basis, we Mellin transform the energy variable $\omega$ via~\eqref{eq:mellin}.
On the celestial sphere, we thus have the following map
\begin{equation}
    \begin{split}
        \omega_i \frac{\pa}{\pa \omega_i} ~\leftrightarrow&~ -\,\Delta_i ~~,~~  t_i\frac{\pa}{\pa t_i} ~\leftrightarrow~ -\,J_i ~~,~~ \omega_i~\leftrightarrow~ e^{\frac{\pa}{\pa\D_i}} ~:=~ T_i
        ~.
    \end{split}
    \label{equ:map}
\end{equation}

\section{Reinterpreting Asymptotic Symmetries}
\label{app:ASG}

In this appendix, we examine more general shifts of the external kinematics, explore their relation to the asymptotic symmetry group, and relate the BCFW shifts to a modified ASG transformation. This expands on the discussion in section~\ref{sec:BCFW} and ties into the analysis in section~\ref{sec:modifiedBCFWtoBG}. 

\subsection{Complex Kinematics and Asymptotic Symmetries}\label{app:zshiftASG}

 Starting with the real, on-shell momenta~\eqref{eq:pi} we introduce a complexified shift of the form \footnote{Throughout this section we will use $z$ as a celestial sphere coordinate and $w$ for our complex shift parameter.}
\be\label{shiftpi}
\hat{p}_i~=~ p_i~+~w\,\epsilon_i\,\o_i\,r_i,~~~w\in \mathbb{C}
\ee
such that the on-shell condition $\hat{p}_i^2=0$ is preserved.  For now we will ignore the additional restriction that the sum over momenta be preserved, which will impose additional constraints on the $r_i$.\footnote{For a discussion of how translation invariance imposes constraints on the support of celestial amplitudes see~\cite{Mizera:2022sln}. }
In order for this shifted momentum to be on-shell for all values of $w$ we must demand
\be\label{ricrit}
r_i^2=p_i\cdot r_i=0
\ee
where $r_i$ can be complex. Writing $p_i=\epsilon_i\omega_i q_i$ in terms of the reference momenta $q_i$, we see that the set
\be\label{equ:refmomenta-basis}
\{q_i,~\p_{z_i} q_i,~\p_{\bz_i} q_i,~\p_{z_i}\p_{\bz_i} q_i\}
\ee
span momentum space. The vector
\be
r_i=\alpha_i q_i+\beta_i\p_{z_i} q_i+\gamma_i\p_{\bz_i} q_i+\delta_i\p_{z_i}\p_{\bz_i} q_i
\ee
meets the criteria~\eqref{ricrit} if
\be
\delta=\beta\gamma=0~.
\ee
We see that there are two options where either $\gamma=0$ or $\b=0$, respectively. For the former, we shift the $i^{\rm th}$ momentum via
\begin{equation}\label{equ:shiftmomentum-1}
    \hat{p}_i~
    ~=~ \epsilon_i\,\o_i\,\Big[\,(1+\a_i w)\,q_i ~+~ \b_i w\, \pa_{z_i}\,q_i\,\Big] 
\end{equation}
which we can also write as a shift of the external data $(\epsilon_i\o_i,z_i,\bz_i)$ 
\begin{equation}
    \begin{split}
        \epsilon_i\o_i ~\mapsto&~ \epsilon_i'\o_i' ~=~ \epsilon_i\o_i\,(1+\alpha_i w)\\
        z_i ~\mapsto&~ z_i'=z_i ~+~ \frac{\beta_i w}{1 + \alpha_i w}\\
        \bz_i ~\mapsto&~ \bz_i ~. \\
    \end{split}
    \label{equ:external-shift-1}
\end{equation}
The second option amounts to the shift
\begin{equation}\label{equ:shiftmomentum-2}
    \hat{p}_i~
    =~ \epsilon_i\,\o_i\,\Big[\,(1+\alpha_i w)\,q_i ~+~ \gamma_i w\, \pa_{\bz_i}\,q_i\,\Big] 
\end{equation}
for which the corresponding shifts in $(\epsilon_i\o_i,z_i,\bz_i)$ are
\begin{equation}
    \begin{split}
        \epsilon_i\o_i ~\mapsto&~ \epsilon_i'\o_i' ~=~ \epsilon_i\o_i\,(1+\alpha_i w)\\
        z_i ~\mapsto&~ z_i \\
        \bz_i ~\mapsto&~ \bz_i'=\bz_i~+~ \frac{\gamma_i w}{1 + \alpha_i w}~~ . \\
    \end{split}
     \label{equ:external-shift-2}
\end{equation}
Now we know that the lightcone in $(2,2)$ is connected and so we expect to be able to identify a Lorentz transformation that takes us from $p_i$ to $\hat{p}_i$.  Comparing to~\eqref{mobius}-\eqref{omegaeps}, we would like to impose
\be
c z_i+d= 1+\alpha_iw,~~~ az_i+b=z_i (1+\alpha_i w)+\beta_i w,~~~ad-bc=1~. \label{zisl2r}
\ee
Since the way we've set up this transformation depends on our starting momenta, the parameters can be $z_i$ dependent.  We find
\be
a=d^{-1}=\frac{1}{1+\alpha_i w},~~~
b=\big(1+\alpha_i w-\frac{1}{1+\alpha_i w}\big)z_i+\beta_iw 
,~~~c=0~.
\ee
Meanwhile the SL$(2,\mathbb{R})_R$ action is trivial $\ba=\bd=1$ and $\bc=\bb=0$. A similar statement holds with the left and right transformations flipped for~\eqref{equ:external-shift-2}.

The class of deformations~\eqref{shiftpi} is a complexification of the on-shell kinematics. By a change of basis we can similarly view this as a complexification of null infinity, and with it the celestial torus/sphere. While the transformation for any particular external particle can be mapped to a (complexified) global  left- or right-handed Lorentz transformation, we want to be able to act on each external particle separately. This is where the asymptotic symmetry group comes in handy. In momentum space the superrotation generators act as follows~\cite{Kapec:2014opa}
\be
\mathcal{L}_{\xi_Y}|in\rangle =\sum_k 
\left[Y^{z}(z_k)\p_{z_k} +\frac{1}{2}(-\omega_k\p_{\omega_k}+s_k)D_zY^z(z_k) \right] |in\rangle
\ee
so that we can induce the desired transformation on each external particle if we match $D_zY^z$ and $Y^z$ at each $z_i$.  For infinitesimal $w$ we have
\be
z_i~\mapsto~ z_i~+~ \beta_i\,w ~-~ \alpha_i\,\beta_i\,w^2 ~+~ \cdots 
\ee
and we want to demand
\be
D_z Y^z(z_i)=-2\alpha_i w~~,~~~Y^z(z_i)=\beta_i w ~.
\ee
This is satisfied by
\be
Y^z=w(\beta_i-2\alpha_i(z-z_i)+\mathcal{O}((z-z_i)^2)+\mathcal{O}(w^2)
\ee
where terms higher order in $(z-z_i)$ and $w$ have been dropped. Now the $z_i$ dependent global Lorentz transformation~\eqref{zisl2r} has vanishing $\mathcal{O}((z-z_i)^2)$ and higher terms. Following \cite{srpi}, we can match this behavior with a meromorphic $Y^z$ of the form 
\begin{equation}\label{small_Y}
    Y^z(z) ~=~ \sum_{i=1}^n\,w\,\Big( A_i +B_i\,(z-z_i) + C_i\, (z-z_i)^2\Big)
    \left(\frac{\prod_{k\neq i}^n(z-z_k)}{\prod_{k\neq i}^n(z_i-z_k)}\right)^3
\end{equation}
where
\begin{align}
    A_i ~=&~ \beta_i \\
    B_i ~=&~ 2\alpha_i-3\beta_i\sum_{j\neq i}\frac{1}{z_i-z_j}  \\
    C_i ~=&~ -3\alpha_i \left[\,2\left(\sum_{k\neq i}^n \frac{1}{z_i-z_k}\right)^2 -\sum_{k>l,k,l\neq i}^n\frac{1}{(z_i-z_k)(z_i-z_l)} \right] .
\end{align}
Now the power of the superrotation Ward identity is that this hard action $Q_H=-i\mathcal{L}_{\xi_Y}$ is equal to a soft insertion
\be
\langle out|[Q,\mathcal{S}]|in\rangle=0~~,~~~~Q=Q_S+Q_H~~ .
\ee
At the linearized level, one can replace this transformation of the external scatterers with a soft insertion. From the celestial perspective, this is equivalent to using the celestial stress tensor to implement such left-handed conformal transformations. We will apply this in section~\ref{sec:modifiedBCFWtoBG} to identify the BG null state starting from a particular generalized BCFW shift we'll discuss in the next section.

\subsection{BCFW and Asymptotic Symmetries}\label{appen:BCFW-ASG}

The interpretation of the BCFW shift as angle-dependent Lorentz transformations discussed in section \ref{sec:celestial-recursion} motivates us to explore their relation to the asymptotic symmetry group (at least at the infinitesimal level).  Starting with the standard BCFW shift (\ref{equ:bcfw-shift}), we can rewrite the shifted external momenta in the basis of (\ref{equ:refmomenta-basis}) as follows\footnote{For simplicity, we set little group factors $t_i=t_j=1$.}
\begin{equation}\label{eq:genshift}
    \begin{split}
        \hat{p}_i ~=&~ \epsilon_i\,\omega_i\,\Bigg[\,\left(1+z\frac{\epsilon_j\sqrt{\omega_j}}{\epsilon_i\sqrt{\omega_i}}\right)\,q_i  ~-~ z\frac{\epsilon_j\sqrt{\omega_j}}{\epsilon_i\sqrt{\omega_i}}z_{ij}\,\p_{z_i}q_i \Bigg] ~~,\\
        \hat{p}_j ~=&~ \epsilon_j\,\omega_j\,\Bigg[\,\left(1-z\frac{\sqrt{\omega_i}}{\sqrt{\omega_j}} \right)\,q_j ~-~ z\frac{\sqrt{\omega_i}}{\sqrt{\omega_j}}\bz_{ij}\,\p_{\bz_j}q_j \Bigg]~.\\
    \end{split}
\end{equation}
Comparing with (\ref{equ:shiftmomentum-1}) - (\ref{equ:external-shift-2}) we can identify the shift variable $z$ as 
\be
z~\mapsto~  w\,\frac{(-2)\omega_i}{\langle ij \rangle}~~,
\ee
so that 
\begin{equation}
    \begin{split}
        \alpha_i ~=&~ \frac{1}{z_{ij}} ~~,~~ \beta_i ~=~ -1 ~~;~~ 
        \alpha_j ~=~ -\frac{1}{z_{ij}}\,T_iT_j^{-1} ~~,~~ \gamma_j ~=~ -\frac{\bz_{ij}}{z_{ij}}\,T_iT_j^{-1} ~
    \end{split}
\end{equation}
while the $\{\alpha_k,\beta_k,\gamma_k\}$ for $k\neq i,j$ are zero. Following the same analysis as in appendix \ref{app:zshiftASG}, the corresponding superrotation killing vectors read
\begin{equation}\label{BCFW_Y}
\begin{split}
    Y^z(z) ~=&~ w\,\Big( A_i +B_i\,(z-z_i) + C_i\, (z-z_i)^2\Big)
    \left(\frac{\prod_{k\neq i}^n(z-z_k)}{\prod_{k\neq i}^n(z_i-z_k)}\right)^3 \\
     Y^{\bz}(\bz) ~=&~ 
     w\,\Big( A_j +B_j\,(\bz-\bz_j) + C_j\, (\bz-\bz_j)^2\Big)
    \left(\frac{\prod_{k\neq j}^n(\bz-\bz_k)}{\prod_{k\neq j}^n(\bz_j-\bz_k)}\right)^3 \\
\end{split}
\end{equation}
where
\begin{align}
    A_i ~=&~ \beta_i ~~,~~ A_j ~=~ \gamma_j ~~, \\
    B_i ~=&~ 2\alpha_i-3\beta_i\sum_{k\neq i}\frac{1}{z_i-z_k} ~~,~~ B_j ~=~ 2\alpha_j-3\beta_j\sum_{k\neq j}\frac{1}{\bz_j-\bz_k}~~,  \\
    C_i ~=&~ -3\alpha_i \left[\,2\left(\sum_{k\neq i}^n \frac{1}{z_i-z_k}\right)^2 -\sum_{k>l,k,l\neq i}^n\frac{1}{(z_i-z_k)(z_i-z_l)} \right] ~~,\\
    C_j ~=&~ -3\alpha_j \left[\,2\left(\sum_{k\neq j}^n \frac{1}{\bz_j-\bz_k}\right)^2 -\sum_{k>l,k,l\neq j}^n\frac{1}{(\bz_j-\bz_k)(\bz_j-\bz_l)} \right]~.
\end{align}
As we have pointed out at the end of section~\ref{sec:celestial-recursion}, the parameters $\a_j$ and $\gamma_j$ are energy-dependent and we treat $\omega_k$ as weight-shifting operator $T_k$.

\section{Helpful Examples}\label{app:examples}
In this appendix, we work through some example computations that expand upon our results in section~\ref{sec:BG} and our discussion in section~\ref{sec:disccussion}.

\subsection{BG PDEs Starting from Parke-Taylor}\label{appen:BGfromParkeTaylor}

In this appendix, we provide a quicker route to the tower of PDEs that extends the BG equations.   Consider a finite BCFW shift~\eqref{eq:hatA} acting on an MHV amplitude.  While the BCFW recursion relations can be used to set up an inductive proof for the $n$-point Parke-Taylor amplitude, for this appendix we will start by assuming we already have this result. Using the Parke-Taylor formula for color-ordered $A[123\cdots n]$, we have
\begin{equation}
    \sum_{p=0}^{\infty}\,\frac{1}{p!}\,z^p\,(D_{ij})^p\, A_n ~=~ \sum_{p=0}^{\infty}\,\sum_{l=0}^p\,(-1)^p\,z^p\,\left(\frac{\langle i-1,j \rangle}{\langle i-1,i \rangle}\right)^{p-l}\left(\frac{\langle i+1,j \rangle}{\langle i+1,i \rangle}\right)^l
    \,A_n[123\cdots n]
    \label{equ:PT-PDE2}
\end{equation}
We can map this to the celestial sphere using \eqref{equ:Dij-CS}.  Comparing both sides of (\ref{equ:PT-PDE2}) at each order of $z$ provides an infinite tower of PDEs for celestial MHV gluon amplitudes.
At $\mathcal{O}(z^p)$ we have\footnote{When deriving (\ref{equ:deriving-OPE}), we noticed that the integrals in the celestial OPE computation only factorized when $t_1=t_2=t$. In what follows, we will set $t_i=t_j$.} \begin{equation}
    \begin{split}
        \Bigg\{\,\Big[\,-\, \Omega_{i,j}\,\left(\,2h_i+z_{ij}\,\pa_i \,\right) +& \Tilde{\Omega}_{i,j}\,\left(\,2\Bar{h}_j+\Bar{z}_{ji}\,\Bar{\pa}_j \,\right)\,\Big]^p \\
        &~-~ (-1)^{p}\,p!\,\Omega^p\,\sum_{l=0}^p\,\left(\frac{z_{i-1,j}}{z_{i-1,i}}\right)^{p-l}\,\left(\frac{z_{i+1,j}}{z_{i+1,i}}\right)^l
        \,\Bigg\}\,\Tilde{A} ~=~ 0
    \end{split}
    \label{equ:k-th-PDEs}
\end{equation}
for $(J_i,J_j)=\{(+,\pm),(-,-)\}$. The analogous computation for $(J_i,J_j)=(-,+)$ will have extra terms but is straightforward to derive. Let us now focus on the case $j=i-1$. In this case only the $\ell=p$ term contributes.  The $p=0$ equation is trivial, while the $p=1$ equation becomes
\begin{equation}
    \begin{split}
       \left(\,\pa_i + \frac{\Delta_i}{z_{i,i-1}} - \frac{1}{z_{i+1,i}} ~+~ \epsilon_i\epsilon_{i-1}\,\frac{2\bar{h}_{i-1}-1+\Bar{z}_{i-1,i}\Bar{\pa}_{i-1}}{z_{i-1,i}}\,T_i\,T^{-1}_{i-1}\,\right) \Tilde{A}~=~0.
    \end{split}
    \label{equ:color-ordered-BG2}
\end{equation}
This is exactly the color-ordered BG equation.

Note that only the $i, i-1$ case is relevant for the celestial OPE. We saw in section~\ref{sec:BG} that for this shift we expect the BG equation exponentiates to the full $p>0$ tower.  Starting from the MHV amplitude here, we see that for the spin configurations for which this shift is allowed, the numerator is annihilated by $D_{i,i-1}$ while relation~\eqref{equ:PT-PDE} follows from the form of the Parke-Taylor denominator.

\subsection{\texorpdfstring{Generalizing to other $D_{i,j}$}{Generalizing to Other Dij}}\label{appen:Dij-PDEs}
The derivation in section~\ref{sec:mhvpde} was for the specific case that $j=i-1$. We will now present a generalization of (\ref{equ:PT-PDE}) to arbitrary $i,j$, still focusing on the MHV case. When $i$ and $j$ are not adjacent, we can draw two non-vanishing channels that give simple poles in $\hat{A}(z)$.  In the MHV case these occur at $z_1 = -\,\frac{\langle i,i+1 \rangle}{\langle j,i+1 \rangle}$ and $z_2 = -\,\frac{\langle i,i-1 \rangle}{\langle j,i-1 \rangle}$. 
In this case (\ref{equ:Taylor-series-A(z)}) still holds while
\begin{equation}
    \begin{split}
        a_p~=&~ \frac{1}{2\pi i}\oint_{\gamma}\,\frac{\hat{A}(z)}{z^{p+1}}\,dz 
    ~=~ \frac{-\,{\rm Res}_{z\to z_1}\, \hat{A}(z)}{z_1^{p+1}}  ~+~ \frac{-\,{\rm Res}_{z\to z_2}\, \hat{A}(z)}{z_2^{p+1}}
    \end{split}
\end{equation}
and 
\begin{equation}
   A ~=~  \hat{A}(0)~=~ \frac{1}{2\pi i}\oint_{\gamma}\,\frac{\hat{A}(z)}{z}\,dz   ~=~ \frac{-\,{\rm Res}_{z\to z_1}\,\hat{A}(z)}{z_1} ~+~ \frac{-\,{\rm Res}_{z\to z_2}\,\hat{A}(z)}{z_2} ~.
\end{equation}
Since we know $\hat{A}(z)$ is a rational function with two simple poles, we can write
\begin{equation}
    \hat{A}(z) ~=~ \frac{f(z)}{(z-z_1)(z-z_2)}
\end{equation}
where $f(z)$ is a polynomial in $z$. A direct calculation tells us
\begin{equation}
    \begin{split}
       a_p ~=&~ -\,\frac{f(z_1)}{(z_1-z_2)\,z_1^{p+1}}~+~\frac{f(z_2)}{(z_1-z_2)\,z_2^{p+1}}\\
       \hat{A}(0)~=&~ -\,\frac{f(z_1)}{(z_1-z_2)\,z_1}~+~\frac{f(z_2)}{(z_1-z_2)\,z_2}.
    \end{split}
\end{equation}
The assumption that $\hat{A}(z)$ goes to zero when $z\to\infty$ further constrains $f(z)$ to be a linear function so we can set $f(z)=az+b$. This gives 
\begin{equation}
    \begin{split}
        a_p ~=&~ \hat{A}(0)\,\left(\frac{a}{b}\frac{-\frac{1}{z_1^{p}}+\frac{1}{z_2^{p}}}{-\frac{1}{z_1}+\frac{1}{z_2}}+\frac{-\frac{1}{z_1^{p+1}}+\frac{1}{z_2^{p+1}}}{-\frac{1}{z_1}+\frac{1}{z_2}}\right) \\
        ~=&~  \hat{A}(0)\,\Bigg[\,\frac{a}{b}\,\sum_{l=0}^{p-1}\,\left(\frac{1}{z_1}\right)^l\,\left(\frac{1}{z_2}\right)^{p-1-l}+ \sum_{l=0}^p\,\left(\frac{1}{z_1}\right)^l\,\left(\frac{1}{z_2}\right)^{p-l}\,\Bigg].
    \end{split}
\end{equation}
Recall that the coefficients $a$ and $b$ can be functions of the spinors $i$ and $j$, just like $z_1$ and $z_2$.  To be consistent with (\ref{eq:hatA}), $a/b$ has to satisfy the following equation
\begin{equation}
   \left(\, D_{i,j} + \frac{a}{b}\,\right)\,\frac{a}{b} ~=~ 0.
\end{equation}
A trivial solution is $a=0$ which corresponds to the Parke-Taylor formula. 
Taking $a=0$ and plugging in the values of $z_1$ and $z_2$ into the expansion, we recover (\ref{equ:PT-PDE2}). Note that the exponentiated form again implies the tower of PDEs in (\ref{equ:PT-PDE2}) is generated by the $p=1$ term.

\subsection{From Soft to Collinear Limits}\label{appen:softtocollinear}

As discussed in section~\ref{sec:BG}, Banerjee and Ghosh originally derived the BG equations by comparing soft and collinear limits. Here we will show that we can work in the other direction.  Starting from the BG equation and the soft expansion~\eqref{equ:soft-currents}, we are able to write the celestial OPE in a way that rearranges descendants in terms of the 2D conformal (rather than full 4D Poincar\'e) symmetry. 

For simplicity, we will again focus on color-ordered gluon amplitudes.  Equation \eqref{equ:soft-currents} tells us that the $k^{\rm th}$ soft gluon theorem takes the form
\begin{equation}
    \begin{split}
        {\rm lim}_{\D_1\to 1-k}&(\D_1-1+k)\,\langle {\cal O}^{\epsilon_1}_{\D_1,+}{\cal O}^{\epsilon_2}_{\D_2,J_2}\cdots{\cal O}^{\epsilon_n}_{\D_n,J_n} \rangle ~=~\frac{(-1)^k}{k!} \,\frac{g\,\epsilon_1^k}{(-2)\,t_1}\\
        & \sum_{q=0}^{\infty}\sum_{l=0}^k\sum_{m=0}^{k-l}\,\begin{pmatrix}
      k\\
      l
      \end{pmatrix}\begin{pmatrix}
      k-l\\
      m
      \end{pmatrix}\,\Big[\epsilon_2\,T_2^{-1}\,(2\bar{h}_2)\Big]^{k-l} \,\frac{(-1)^m}{z_{n2}^{q+m+l}}\,z_{12}^{q+m+l-1} \\
      &\qquad \Big[\epsilon_n\,T_n^{-1}\,(2\bar{h}_n-\bz_{2n}\bar{\pa}_n)\Big]^{l}\,
        \langle {\cal O}^{\epsilon_2}_{\D_2,J_2}\cdots{\cal O}^{\epsilon_n}_{\D_n,J_n} \rangle
    \end{split}
    \label{equ:k-soft-gluon-theorem}
\end{equation}
where we have considered a generalized collinear limit where we take  $\bz_{12}=0$ and series expand around at $z_{12}\to 0$.  For reference, the first two orders of this expansion take the form
\begin{equation}
    \begin{split}
       & {\rm lim}_{\D_1\to 1-k}(\D_1-1+k)\,\langle {\cal O}^{\epsilon_1}_{\D_1,+}{\cal O}^{\epsilon_2}_{\D_2,J_2}\cdots{\cal O}^{\epsilon_n}_{\D_n,J_n} \rangle ~=~\frac{(-1)^k}{k!} \,\frac{g}{(-2)\,t_1}\,\epsilon_1^k \\
       &\Bigg\{\,\frac{1}{z_{12}} 
       ~+~ \Bigg[\, \frac{(1-k)}{z_{n2}}\Big[\epsilon_2\,T_2^{-1}(2\bar{h}_2)\Big]^k + \frac{k}{z_{n2}}\,\Big[\epsilon_2\,T_2^{-1}(2\bar{h}_2)\Big]^{k-1}\,\Big[\epsilon_n T_n^{-1}(2\bar{h}_n-\bz_{2n}\bar{\pa}_n\Big]\,\Bigg]\\
       &\qquad\qquad\qquad\qquad\qquad\qquad\qquad\qquad\qquad + {\cal O}(z_{12})\,\Bigg\}\langle {\cal O}^{\epsilon_2}_{\D_2,J_2}\cdots{\cal O}^{\epsilon_n}_{\D_n,J_n} \rangle.
    \end{split}
    \label{equ:soft_collinear}
\end{equation}
We now want to compare this with the collinear limits at generic conformal dimensions. Comparing (\ref{equ:color-ordered-BG}) with (\ref{equ:soft_collinear}), we can write the subleading terms as SL$(2,\mathbb{R})_L$ descendants and get the following constraint for the subleading order ${\cal O}(z_{12}^0)$ of the ${\cal O}_{\D_1,+}(z_1,\bz_2){\cal O}_{\D_2,+}(z_2,\bz_2)$ OPE.\footnote{This is based on the fact that the differential equations contain the soft term (\ref{equ:k-soft-gluon-theorem}). For the example we described in the text, (\ref{equ:color-ordered-BG}) contains the soft term at order ${\cal O}(z_{12}^0)$ in (\ref{equ:soft_collinear})}
\begin{equation}
\begin{split}
    &{\rm lim}_{\D_1\to 1-k}(\D_1-1+k)\, \langle {\cal O}^{\epsilon_1}_{\D_1,+}(z_1,\bz_2){\cal O}^{\epsilon_2}_{\D_2,+}(z_2,\bz_2)\cdots{\cal O}^{\epsilon_n}_{\D_n,J_n} \rangle  ~\overset{{\rm subleading}}{=}~ \frac{(-1)^{k+1}}{k!}\,\frac{g}{t_1(-2)}\\
    &~ \epsilon_1^k\,\epsilon_2^k\, \left(\prod_{l=1}^{k-1}(2\bar{h}_2-l)\right)\left(k\,\pa_2 + \frac{\D_2-1}{z_{2n}} + \frac{k}{z_{23}} \right)
    \,T_2^{-k}\,\langle {\cal O}^{\epsilon_2}_{\D_2,+}(z_2,\bz_2)\cdots{\cal O}^{\epsilon_n}_{\D_n,J_n} \rangle
\end{split}
\label{equ:subleading-OPE-constraint}
\end{equation}
where $k=\mathbb{Z}_+$. Now we expect the celestial amplitude to be analytic in the conformal dimension~\cite{Arkani-Hamed:2020gyp}, with poles for the external dimensions restricted to the values at which we've extracted the residues here.\footnote{At loop order one can develop higher-order poles for these values of the conformal dimension, whose residues get a dependence on the IR cutoff.} Based on the RHS of (\ref{equ:subleading-OPE-constraint}), we can write down the following ansatz for the subleading ${\cal O}(z_{12}^0)$ term in the ${\cal O}_{\D_1,+}(z_1,\bz_2){\cal O}_{\D_2,+}(z_2,\bz_2)$ OPE for generic conformal dimensions
\begin{equation}
\begin{split}
    \langle {\cal O}^{\epsilon_1}_{\D_1,+}(z_1,\bz_2){\cal O}^{\epsilon_2}_{\D_2,+}&(z_2,\bz_2)\cdots{\cal O}^{\epsilon_n}_{\D_n,J_n} \rangle ~=~ \frac{g}{t_1(-2)}\,C(\D_1,\D_2)\\
    &~\Bigg\{\,f(\D_1)\,\pa_2 ~+~ \frac{\D_2-1}{z_{2n}} ~+~ \frac{f(\D_1)}{z_{23}}\,\Bigg\}\,\langle {\cal O}^{\epsilon_2}_{\D_1+\D_2-1,+}(z_2,\bz_2)\cdots{\cal O}^{\epsilon_n}_{\D_n,J_n} \rangle\\
    &~~+~~\text{non-singular terms for }\D_1  ~.
    \end{split}
\end{equation} 
Namely, we've treated the OPE as a complex function of $\D_1$.  Whenever the parameter $k$ occurs we replace it with a function of $\D_1$ to be fixed. The poles and residues on the $\D_1$-complex-plane are given, and  our goal is to determine $C(\D_1,\D_2)$ and $f(\D_1)$ so that match onto~\eqref{equ:subleading-OPE-constraint} at the corresponding integer values of the conformal dimension. The non-singular terms cannot be fixed by this discussion and we will see the non-singular terms play an important role when $\epsilon_1=-\epsilon_2$ shortly.  This procedure gives the following constraints
\begin{enumerate}
    \item When $\epsilon_1=\epsilon_2$,
    \begin{equation}\scalemath{1}{
        \badat{3} {\rm Res}_{\D_1\to 1-k}\,C(\D_1,\D_2) &=~ \frac{(-1)^{k+1}}{k!}\,\frac{\Gamma(\D_2-1)}{\Gamma(\D_2-k)}\\
        \Rightarrow~~ C(\D_1,\D_2) &=~ -\,\frac{B(\D_1-1,\D_2-1)}{\D_1+\D_2-2} ~.
        \eadat}
    \end{equation}
    \item When $\epsilon_1=-\epsilon_2$,
    \begin{equation}
    \scalemath{.96}{
        \begin{aligned}
            {\rm Res}_{\D_1\to 1-k}\,C(\D_1,\D_2) &=~ \frac{(-1)^{2k+1}}{k!}\,\frac{\Gamma(\D_2-1)}{\Gamma(\D_2-k)} ~=~ \frac{(-1)^{k}}{k!}\,\frac{\Gamma(\D_2-1)\Gamma(k+1-\D_2)}{\Gamma(\D_2+1)\Gamma(-\D_2)} \\
            ~~~\Rightarrow~~~ C(\D_1,\D_2) &=~ -\,\frac{B(\D_1-1,3-\D_1-\D_2)}{\D_1+\D_2-2} ~.
        \end{aligned}}
    \end{equation}
    \end{enumerate}
   Note that $\D_1$ soft limit in fact only contributes to the 
${\cal O}^{\epsilon}_{\D_1+\D_2-1,+}(z_2,\bz_2)$
sector. In order to consider the ${\cal O}^{-\epsilon}_{\D_1+\D_2-1,+}$ sector, one could consider the $\D_2$ soft limit and the similar calculation yields a prefactor
\begin{equation}
    C(\D_2,\D_1) ~=~ -\,\frac{B(\D_2-1,3-\D_1-\D_2)}{\D_1+\D_2-2} ~.
\end{equation}
Meanwhile, the function $f(\Delta_1)$ is regular at these values of the conformal dimension. Assuming there are no additional poles for finite $\Delta_1$ we can promote this to
    \begin{equation}
         f(1-k)~=~ k ~~~\Rightarrow~~~ f(\D_1) ~=~ 1-\D_1~.
    \end{equation}
Putting this together, we see that for $\epsilon_1=\epsilon_2=\epsilon$ we have
\begin{equation}
\begin{split}
    \langle {\cal O}^{\epsilon}_{\D_1,+}(z_1,&\bz_2){\cal O}^{\epsilon}_{\D_2,+}(z_2,\bz_2)\cdots{\cal O}^{\epsilon_n}_{\D_n,J_n} \rangle ~=~ \frac{g\,B(\D_1-1,\D_2-1)}{t_1(-2)}\,\Bigg\{\,\frac{\D_1-1}{\D_1+\D_2-2}\,\pa_2\\
    &~+~ \frac{\D_2-1}{\D_1+\D_2-2}\frac{1}{z_{n2}} ~+~ \frac{\D_1-1}{\D_1+\D_2-2}\,\frac{1}{z_{23}}\,\Bigg\}\,\langle {\cal O}^{\epsilon}_{\D_1+\D_2-1,+}(z_2,\bz_2)\cdots{\cal O}^{\epsilon_n}_{\D_n,J_n} \rangle~~;\\
        \end{split}
\end{equation}
while  $-\epsilon_1=\epsilon_2=\epsilon$ we have
\begin{equation}
\begin{split}
    \langle {\cal O}^{-\epsilon}_{\D_1,+}(z_1,&\bz_2){\cal O}^{\epsilon}_{\D_2,+}(z_2,\bz_2)\cdots{\cal O}^{\epsilon_n}_{\D_n,J_n} \rangle ~=~ \frac{g\,B(\D_1-1,3-\D_1-\D_2)}{t_1(-2)}\,\Bigg\{\,\frac{\D_1-1}{\D_1+\D_2-2}\,\pa_2\\
    &~+~ \frac{\D_2-1}{\D_1+\D_2-2}\frac{1}{z_{n2}} ~+~ \frac{\D_1-1}{\D_1+\D_2-2}\,\frac{1}{z_{23}}\,\Bigg\}\,\langle {\cal O}^{\epsilon}_{\D_1+\D_2-1,+}(z_2,\bz_2)\cdots{\cal O}^{\epsilon_n}_{\D_n,J_n} \rangle\\
    &~+~ \text{a similar term for } \langle {\cal O}^{-\epsilon}_{\D_1+\D_2-1,+}(z_2,\bz_2)\cdots{\cal O}^{\epsilon_n}_{\D_n,J_n} \rangle.
    \end{split}
\end{equation}
The first equation coincides with the subleading results presented in \cite{Ebert:2020nqf}. 
One can extend this analysis to all orders in $z_{12}$, reproducing the SL($2,\mathbb{R})_L$ block.

\subsection{Revisiting the 4-Gluon Correlator}\label{appen:4gluon}
In this section, we revisit the BCFW construction of the 4-point celestial gluon amplitude in $(2,2)$ signature from~\cite{Pasterski:2017ylz}, using the reformulation in terms of weight shifting operators from~\cite{Guevara:2019ypd}.  This example helps contrast the conformal block and MHV decompositions, so it is useful to see how the celestial BCFW procedure works in this case.

\paragraph{4-Gluon Correlator via (\ref{equ:c-recursion})}
Consider the color-ordered MHV gluon celestial amplitudes where gluons 2 and 3 have negative helicity. In this case, the collinear channel (\ref{equ:c-recursion}) gives the complete $n$-point result $\tilde{A}_n = \tilde{A}_n^c$, and there is only one term for which $s_1=1$, $i=2$, and $s_I=-1$
\begin{equation}
    \begin{split}
      \tilde{A}^{+--+ \cdots +}_n ~=&~ g\,\frac{z_{n2}}{(-2)z_{12}z_{n1}t_1}\,\int_0^{\infty}\frac{d\o_1}{\o_1}\,\o_1^{\D_1}\,\frac{1}{\o_1}\,\left|1+\a \right|^{-2\bar{h}_2}\,\left|1+\b \right|^{-2\bar{h}_n}\\
       & \tilde{A}^{--+\cdots +}_{n-1}\left(\epsilon_2{\rm sgn}(1+\a),z_2,\bar{z}_2+\frac{\a\bar{z}_{12}}{1+\a};\dots;\epsilon_n{\rm sgn}(1+\b),z_n,\bar{z}_n+\frac{\b\bar{z}_{1n}}{1+\b} \right) 
    \end{split}
    \label{equ:recursion-MHVgluon}
\end{equation}
where
\be
     \a ~=~ \o_1\,\frac{\epsilon_1 z_{n1}}{\epsilon_2 z_{n2}}\,T_2^{-1}~~~,~~~\b ~=~ \o_1\,\frac{\epsilon_1 z_{21}}{\epsilon_n z_{2n}}\,T_n^{-1}.
\ee
We will now restore the momentum conserving delta functions. First, the 3-point function is \cite{Pasterski:2017ylz}
\begin{equation}
    \tilde{A}_3^{--+} ~=~ (-\pi)\,\d\left(\sum_{i=1}^3\D_i-3\right)\,\frac{{\rm sgn}(z_{12}z_{23}z_{34})\d(\bar{z}_{12})\d(\bar{z}_{23})}{|z_{12}|^{-\D_3}|z_{23}|^{2-\D_1}|z_{31}|^{2-\D_2}}\,\Theta\left(\frac{\epsilon_2z_{12}}{\epsilon_3z_{31}}\right)\,\Theta\left(\frac{\epsilon_2z_{23}}{\epsilon_1z_{31}}\right).
     \label{equ:3pt-MHVgluon}
\end{equation}
Together with~\eqref{equ:recursion-MHVgluon} this implies the following expression at 4-point
\begin{equation}\label{equ:4pt-MHVgluon-step1}
    \scalemath{0.94}{\badat{3}
      \tilde{A}^{+--+}_4 ~=&~ \frac{g}{t_1}\,\frac{\pi}{2}\,{\rm sgn}(z_{12}z_{23}z_{34}z_{41})\,\frac{|z_{42}|}{|z_{12}z_{41}|}\,\int_0^{\infty}\,d\o_1\,\o_1^{\D_1-2}\,\left|1+\a \right|^{-1-\D_2}\,\left|1+\b \right|^{1-\D_4}\\
       &\frac{\d\left(\sum_{i=2}^4\D_i-3\right)}{|z_{23}|^{-\D_4}|z_{34}|^{2-\D_2}|z_{42}|^{2-\D_3}}\,\d\left(\bar{z}_{13}-\frac{\bar{z}_{12}}{1+\a}\right)\,\d\left(\bar{z}_{13}-\frac{\bar{z}_{14}}{1+\b}\right)\,\Theta\left(\frac{\epsilon_3z_{23}}{\epsilon_4'z_{42}}\right)\,\Theta\left(\frac{\epsilon_3z_{34}}{\epsilon_2'z_{42}}\right). 
  \eadat}
\end{equation}
First, let us compare this to the result in~\cite{Pasterski:2017ylz}. If we were starting from the 4-point kinematics we would expect that three delta functions localizing all three independent energy ratios $\omega_i/\omega_j$, one delta function equating the left and right cross ratios, and the remaining $\o$-integral yields an additional delta function forcing $\sum_i\Delta_i=4$.  We see this from the three-point kinematics in~\eqref{equ:4pt-MHVgluon-step1} as follows:
\begin{enumerate}
    \item The first delta function can be used to localize the $\omega_1$ integral.  This leave us with a power of the weight shifting operator $T_2^{\D_1-1}$: $\D_2$ $\to$ $\D_2+\D_1-1$.  This has the effect of remedying the delta function support for the weights
    \begin{equation}
        \d\left(\sum_{i=2}^4\D_i-3\right) ~\to~ \d\left(\sum_{i=1}^4\D_i-4\right)
    \end{equation}
    to match what we expect at 4-point.
    \item The second delta function enforces $T_2T^{-1}_4
    =\frac{\epsilon_4\bar{z}_{34}{z}_{14}}{\epsilon_2\bar z_{23}{z}_{12}}$ however there is no extra integration occurring.  Because $T_2T^{-1}_4$ should be invariant under a reflection that swaps $z_{ij}$ and $\bz_{ij}$ we can rewrite this delta function as 
    \begin{equation}
        \d\left(\bar{z}_{43}-\frac{z_{21}z_{34}\bar{z}_{14}\bar{z}_{32}}{z_{41}z_{23}\bar{z}_{12}}\right) ~=~ \left|\frac{(1-z)z}{\bar{z}_{34}}\right|\,\d(z-\bar{z})
    \end{equation}
    which is the analytic continuation to (2,2) of the reality condition on the cross-ratio for massless kinematics in (1,3).
\end{enumerate}
After some algebra, we finally get
\begin{equation}
    \begin{split}
         \tilde{A}^{+--+}_4 ~=&~ \frac{g}{t_1}\,\frac{\pi}{2}\,{\rm sgn}(z_{12}z_{23}z_{34}z_{41})\,\d\left(\sum_{i=1}^4\D_i-4\right)\left(\prod_{i<j}\,|z_{ij}|^{\frac{h}{3}-h_i-h_j}\,|\bar{z}_{ij}|^{\frac{\bar{h}}{3}-\bar{h}_i-\bar{h}_j}\right)\\
         & \d(z-\bar{z})\,|z|^{-\frac{1}{3}}|1-z|^{\frac{5}{3}}\,
         \Theta\left(\frac{\epsilon_3z_{23}\bar{z}_{13}}{\epsilon_4z_{42}\bar{z}_{14}}\right)\,
         \Theta\left(\frac{\epsilon_3z_{34}\bar{z}_{13}}{\epsilon_2z_{24}\bar{z}_{12}}\right)\,
         \Theta\left(\frac{\epsilon_2z_{42}\bar{z}_{32}}{\epsilon_1z_{41}\bar{z}_{13}}\right) 
    \end{split}
     \label{equ:4pt-MHVgluon}
\end{equation}
where $h=\sum_i h_i = 2$ and $\bar{h}=\sum_i \bar{h}_i = 2$. This matches what we get from directly Mellin transforming the momentum space 4-point MHV gluon amplitude in (2,2) signature.

\paragraph{Gluing 3-pt Correlators}
We can now recap the computation in~\cite{Pasterski:2017ylz} as follows. A $[1,4\rangle$ shift of the form~\eqref{equ:bcfw-shift} is little group equivalent to the following shifted kinematics
\begin{align}
\begin{split}\label{hatted}
& (\hat\epsilon_1\hat \omega_1 , \hat z_1 , \hat {\bar z}_1 ) = \left( (1-z \zeta ) \epsilon_1\omega_1 ,z_1 , \frac{\bar z_1 -z\zeta \bar z_4}{ 1-z\zeta}\right)\,,\\
&(\hat \epsilon_2\hat \omega_2  , \hat z_2 , \hat {\bar z}_2) =(\epsilon_2 \omega_2, z_2,\bar z_2)\,,\\
&(\hat \epsilon_3\hat \omega_3  , \hat z_3, \hat {\bar z}_3 ) = (\epsilon_3\omega_3 ,z_3 , \bar z_3)\,,\\
&(\hat \epsilon_4\hat \omega_4  , \hat z_4, \hat {\bar z}_4 )  =
\left( (1-z \zeta^{-1})\epsilon_4 \omega_4 ,\frac{ z_4 -z\zeta^{-1} z_1}{ 1-z\zeta^{-1}} ,\bar z_4\right)\,,
\end{split}
\end{align}
where
\be
\zeta=\sqrt{\frac{\omega_4}{\omega_1}}
\ee
and for reference $U\leftrightarrow-z$ in the notation of~\cite{Pasterski:2017ylz}. 
By incorporating the covariant $\epsilon$'s we can drop the absolute values there.
 While the momentum conservation locus is preserved and each of the external momenta is on-shell, the intermediate momenta obeys
\be
P^2 = \hat{P}_{12}^2=P_{12}^2+ 4z\zeta\omega_1\omega_2 z_{12}\bz_{24}~~,~~
\epsilon_P\,\o_P = \epsilon_1\,\o_1 + \epsilon_2\,\o_2 -z\,\epsilon_1\,\o_1\,\zeta~.
\ee
For complex $z$ these acquire imaginary parts.  In $(2,2)$ signature, however, we see that the pole is on the real $z$ axis and the fact that $\zeta$ is real means that the shifted celestial coordinates will also be real.

The BCFW expression for the 4-gluon amplitude is written out above. As we saw, there is only one (collinear) factorization channel and
 $A^c_4$ splits off one anti-MHV 3-point function by construction, and leaves the answer in terms of another $n-1=3$ point function.  All we are aiming to do is to rewrite it in a form more reminiscent of a recursion between lower point unstripped amplitudes. As shown in~\cite{Pasterski:2017ylz} we have
\begin{align}
&\mathcal{\tilde A}_{--++}(\lambda_i , z_i ,\bar z_i ) \notag \\
&~\propto
\int_{-\infty}^\infty \frac{dz}{|z|}   \,     
   \int_{-\infty}^\infty  \frac{d \lambda_P}{ 2\pi}  
   \int d z_P d{\bar z}_P
\mathcal{\tilde A}_{--+}( \lambda_1,\lambda_2 , \lambda_P;  \hat z_j , \hat{\bar z}_j)
\mathcal{\tilde A}_{-++}( -\lambda_{P}, \lambda_3, \lambda_4 ; \hat z_j , \hat{\bar z}_j)
\,.
\label{BCFWMellin}
\end{align}
where $z$ is now real, the hatted variables are defined in \eqref{hatted}, and $\hat{z}_p=z_p$, $\hat{\bz}_p=\bz_p$. Here we've dropped some kinematical factors to focus on the integrated quantities.

\paragraph{BCFW vs Conformal Block}
We now want to compare the BCFW expression (\ref{BCFWMellin}) to what we would expect from a conformal block expansion for the 4-point case.   First, we notice that our discussion above offers a reinterpretation of the $z$-integral in~\cite{Pasterski:2017ylz}.
Although the contour of integration is different, it plays the same role as the BCFW residue.  In the standard recursion relations, the pole in $z$  picks out a configuration where the amplitude goes on-shell.  Here it is saturated on the locus  $\delta(P^2)$  similarly enforcing the intermediate momentum being on-shell.  By contrast, we note that if we examined the $s$-channel perturbative amplitude rather than the on-shell recursion, the split representation of the propagator~\cite{Melton:2021kkz} would involve a similar four-dimensional integral but now include all values of the intermediate $P^2$. Meanwhile, the smearing over $\lambda_P$, $z_p$, and $\bz_P$ looks like a slight modification of the projectors appearing in conformal blocks~\cite{SimmonsDuffin:2012uy}, perhaps unsurprisingly due to the non-standard 2-point functions unless we introduce additional shadow transforms~\cite{Pasterski:2017kqt,ss,Crawley:2021ivb}.

\bibliographystyle{utphys}
\bibliography{references}

\end{document}